\RequirePackage{fix-cm}
\documentclass[twocolumn]{svjour3}          

\smartqed  
\usepackage{graphicx}
\usepackage{booktabs}

\usepackage{color}
\usepackage{url}		
\usepackage{tabularx}
\usepackage{float}

 \usepackage{multirow}

\usepackage{amsmath}
\usepackage{nccmath}

\usepackage{pdfpages}

\usepackage[toc,page]{appendix}

\usepackage{listings}
\usepackage{times}
\AtBeginDocument{}

\definecolor{gray97}{gray}{.97}
\definecolor{gray75}{gray}{.75}
\definecolor{gray45}{gray}{.45}

\lstnewenvironment{listing}[1][]
{\lstset{#1}\pagebreak[0]}{\pagebreak[0]}

\lstdefinestyle{consola}
{basicstyle=\scriptsize\bf\ttfamily,
	backgroundcolor=\color{gray75},
}

\lstdefinestyle{JAVA}
{language=JAVA,
}

\begin{document}

\title{Leveraging creativity in requirements elicitation within agile software development: a systematic literature review}

\author{Ainhoa Aldave \and Juan M. Vara \and David Granada \and Esperanza Marcos }

\institute{
	A. Aldave, J.M. Vara, D. Granada, E. Marcos \at Kybele Research Group, Rey Juan Carlos University,\\	Calle Tulip\'an S/N, 28933, M\'ostoles, Madrid, Spain           
	\and
	A. Aldave \at
	\email{a.aparicioa@alumnos.urjc.es} 
	\and
	J.M. Vara \at
	\email{juanmanuel.vara@urjc.es}
	\and
	D. Granada \at
	\email{david.granada@urjc.es}
	\and
	E. Marcos \at
	\email{esperanza.marcos@urjc.es}
}

\date{Received: date / Accepted: date}

\maketitle

\begin{abstract}
Agile approaches tend to focus solely on scoping and simplicity rather than on problem solving and discovery. This hampers the development of innovative solutions. Additionally, little has been said about how to capture and represent the real user needs. To fill this gap, some authors argue in favor of the application of ``Creative thinking'' for requirements elicitation within agile software development. This synergy between creativeness and agility has arisen as a new means of bringing innovation and flexibility to increasingly demanding software. 

The aim of the present study is therefore to employ a systematic review to investigate the state-of-the-art of those approaches that leverage creativity in requirements elicitation within Agile Software Development, as well as the benefits, limitations and strength of evidence of these approaches. 

The review was carried out by following the guidelines proposed by Dr. Kitchenham. The search strategy identified 1451 studies, 17 of which were eventually classified as primary studies.  The selected studies contained 13 different and unique proposals. These approaches provide evidence that enhanced creativity in requirements elicitation can be successfully implemented in real software projects. We specifically observed that projects related to user interface development, such as those for mobile or web applications, are good candidates for the use of these approaches. We have also found that agile methodologies such as Scrum, Extreme Programming or methodologies based on rapid modelling are preferred when introducing creativity into requirements elicitation. Despite this being a new research field, there is a mixture of techniques, tools and processes that have already been and are currently being successfully tested in industry. Finally, we have found that, although creativity is an important ingredient with which to bring about innovation, it is not always sufficient to generate new requirements because this needs to be followed by user engagement and a specific context in which proper conditions, such as flexibility, time or resources, have to be met.

\keywords{Software Development \and Software Project Management \and Agile Methodologies \and Requirements Elicitation \and Creative Thinking \and Systematic Review}
\end{abstract}

\section{Introduction}
\label{sec:introduction}
\sloppy
In today's software development environments, in which needs, technology, complexity and demands evolve so quickly, the need for innovation and agility has become crucial and very often the reason for the disruptive gap between success and failure \cite{lehman1980programs,lehman1997metrics}. Software ideas need to be functional as soon as possible and sufficiently original to beat competition and meet the expectations of their increasingly demanding users. 

\sloppy
Agile methodologies such as Extreme Programming, Kanban or Scrum have, therefore, emerged to deal with the increasing complexity in software engineering and to handle the inevitable changes in requirements throughout the life-cycle of software. Agile software development methodologies have already proven their effectiveness as regards delivering software more quickly, improving customer collaboration, estimating time and handling defects in processes \cite{dybaa2008empirical}. While agility is already established in organizations of all sorts, the time has come to confront its limitations.

Some studies suggest that the existing agile approaches employed during the requirements analysis phase tend to focus solely on scoping and simplicity rather than on problem solving and discovery \cite{turk2014limitations,hosbond2008misfit}. This results in the inability to develop innovative solutions. Additionally, even though it is claimed that agile meth\-od\-o\-lo\-gies provide collaboration between the customer and the development team, very few define how the people on the team playing the role of the customer can learn what the real end user needs and how they can accurately represent those needs [S17].

If agility is apparently not a sufficient means to provide innovative solutions, how can we enhance innovation in software development? Innovation has been defined as the successful implementation of creative ideas \cite{hennessey1988conditions}, which leads us to the question: how can we successfully enhance creativity in software development? 

Within the field of Design, creativity has been widely considered as an essential prerequisite to enable the emergence of innovative ideas in product development \cite{snider2011studying}. Some creative approaches, such as Design Thinking, have suggested models and strategies whose objective is to facilitate the delivery of innovative ideas \cite{lindberg2011design,brown2008design,martin2009design}. But, why introduce creativity only into Design? Would it be possible to apply similar approaches to other phases of software development? 

Software development usually starts with requirements analysis, which plays a determining role in the development process and is still considered as one of the most critical activities in any software development project \cite{davis2006effectiveness}. In fact, we have discovered that, during the last decade, engineers have intensively started to underline the importance of creative thinking in requirements elicitation (RE) as a decisive factor for building competitive and innovative products \cite{maiden2001requirements,maiden2010requirements}.

Several systematic reviews on the application of agile meth\-od\-o\-lo\-gies in the requirements engineering currently exist, each of them focused on different aspects of the process of defining, documenting and maintaining requirements \cite{curcio2018requirements,inayat2015systematic,schon2017agile,heikkila2015mapping,salah2014systematic}. We have also found some works studying the state-of the art regarding creativity in agile systems development \cite{conboy2009creativity} or creativity in requirements engineering \cite{lemos2012systematic}. However, we have not, to date, been able to find a systematic review focused on combining both creative approaches for requirements elicitation within agile software development. This paper aims to fill this gap by presenting a systematic review of the existing work in this field.

Likewise, Yang et al. \cite{yang2016systematic}, present a detailed analysis of the literature on the combination of software architecture and agile development, in which they reveal a series of reflections and implications about this combination. These results can be useful to the community of software engineers in order to select appropriate architecture activities, approaches and agile practices. We consider that this type of studies, with their respective conclusions, are complementary to the results obtained in this paper, in which we add the factor of creativity, a concept that has aroused considerable interest in recent years in software engineering community.

The main goal of this paper is to study the state-of the art of the approaches that strengthen creativity in requirements elicitation within agile software development. 

The following sub-goals have, therefore, been defined: 

\begin{itemize}
\item The research on background studies: requirements elicitation, agile methodologies and creativity.
\item The definition of a criterion that will serve to select and evaluate relevant studies. 
\item The execution of a systematic literature review.
\item The review and summary of the selected studies and the identification of trends. 
\end{itemize}

Before presenting this review, it is worth clarifying that, although not all the analyzed studies contain empirical evidence, we understand that the object of study is specific enough to classify this study as a systematic review, understanding that within the historical debate between systematic mappings and systematic literature reviews, it is almost always feasible to find reasons to typify a systematic analysis of the literature in one way or another. 

Furthermore, in this case we have chosen to include an analysis of the quality of the works (a task that is not essential in a mapping study, but it is in a systematic review) and we have included an ad-hoc research question to identify the empirical evidence provided by each primary study selected. However, given that the review protocol is essentially the same, we understand that the typification of the work in one way or another does not influence or detract from the scientific contribution.

The remainder the paper is organized as follows: Section \ref{sec:researchContext} describes the background that was studied prior to carring out this work: requirements elicitation, agile methodologies and creativity; Section \ref{sec:SLRProcess} explains the method adopted to conduct the systematic review; Section \ref{sec:dataExtraction} provides a detailed review of each of the results of the systematic review; Section \ref{sec:dataSynthesisResults} summarizes the findings and provides a discussion, and finally, Section \ref{sec:conclusions} summarizes the main conclusions derived from this paper and outlines our future work.

\section{Background studies}
\label{sec:researchContext}
This paper aims to evaluate existing creative approaches for requirements elicitation within the context of agile software development. This section therefore introduces the main concepts that must be understood in the context of the present study: Requirements Elicitation, Agile Methodologies and Creativity.

\subsection{Requirements Elicitation}
\label{sec:ReqElic}

Requirements elicitation is the process of defining stakeholders' needs and putting this information together in an understandable manner such that developers can construct a system that will address those needs \cite{tiwari2012selecting}.

Although requirements elicitation is already a relatively mature area in the software development industry \cite{hickey2004unified} it is still considered to be one of the most critical activities in any software development project \cite{davis2006effectiveness}.

One of the main challenges of requirements elicitation is that of fully understanding what the users really want. Interaction with the users usually takes place via natural language, which is not always straightforward. The consequence is that users tend to provide incomplete or ambiguous requirements. Another problem is that the requirements gathered during the early stages of the project are likely to evolve or be discarded in later phases of the project. A lack of user involvement or having unrealistic expectations are also common problems. Since the social context is much more crucial than the technical one, these issues cannot be solved in a purely technological way \cite{goguen1993techniques}.

Requirements elicitation techniques have been studied in depth in a wide number of studies \cite{tiwari2012selecting,davis2006effectiveness,hickey2003elicitation,goguen1993techniques,zowghi2005requirements,maiden1996acre}. Since the number of different techniques and approaches is extensive, there is no standard categorization. Some of the most popular include: interviews, questionnaires, surveys, scenarios, brainstorming, user stories, workshops, role playing and prototyping. Within the context of this paper, we focus principally on requirements elicitation techniques and approaches that promote creativity and user engagement.

\subsection{Agile Methodologies}
\label{sec:AgileMet}

Within the field of software development, in which needs, technology, complexity and demands evolve so quickly \cite{lehman1980programs,jiang2009analysis}, the need for innovation and agility has become crucial and very often the reason for the disruptive gap between success and failure. Agile methodologies, such as the Dynamic Software Development Method \cite{stapleton2003dsdm}, Extreme Programming \cite{beck2000extreme}, Crystal \cite{cockburn2004crystal}, Feature-driven development \cite{palmer2001practical}, Kanban \cite{anderson2010kanban} or Scrum \cite{schwaber2016scrum}, have emerged to deal with the increasing complexity in software engineering and to handle the inevitable changes in requirements throughout their life-cycle \cite{jiang2009analysis}.

In 2001, a group of 17 independent-minded software consultants and practitioners gathered together and signed the Agile Software Development Manifesto, which promulgated the following core values and principles: 

\begin{itemize}

\item Values:
\begin{itemize}
\item[$\cdot$] Individuals and interactions over processes and tools
\item[$\cdot$] Working software over comprehensive documentation
\item[$\cdot$] Customer collaboration over contract negotiation
\item[$\cdot$] Responding to change over following a plan 
\end{itemize}

\item Principles:
\begin{itemize}
\sloppy
\item[$\cdot$] The highest priority is to satisfy the customer through the early and continuous delivery of valuable software
\item[$\cdot$] Welcome changing requirements, even late the development. Agile processes harness change for the customer's competitive advantage
\item[$\cdot$] Deliver working software frequently, from a couple of weeks to a couple of months, with a preference for the shorter timescale
\item[$\cdot$] Business people and developers must work together daily throughout the project
\item[$\cdot$] Build projects around motivated individuals. Give them the environment and support they need and trust them to get the job done
\item[$\cdot$] The most efficient and effective method by which to convey information to and within a development team is face-to-face conversation
\item[$\cdot$] Working software is the primary measure of progress
\item[$\cdot$] Agile processes promote sustainable development. The sponsors, developers, and users should be able to maintain a constant pace indefinitely
\item[$\cdot$] Paying continuous attention to technical excellence and good design enhances agility
\item[$\cdot$] Simplicity -the art of maximizing the amount of work not done- is essential
\item[$\cdot$] The best architectures, requirements and designs emerge from self-organizing teams
\item[$\cdot$] At regular intervals, the team should reflect on how to become more effective, then tune and adjust its behavior accordingly
\end{itemize}
\end{itemize}

Agile Methodologies have already been proven to be beneficial as regards reducing the delivery time of working software, improving customer collaboration, estimating time and handling defect processes \cite{dybaa2008empirical}.

In the following subsections, we briefly describe some of the most popular Agile Methodologies: Extreme Programming, Kanban and Scrum.

\subsubsection{Extreme Programming}
\label{sec:ExtremeProg}

\begin{figure}[H]
	\centering
	\includegraphics[width=\columnwidth]{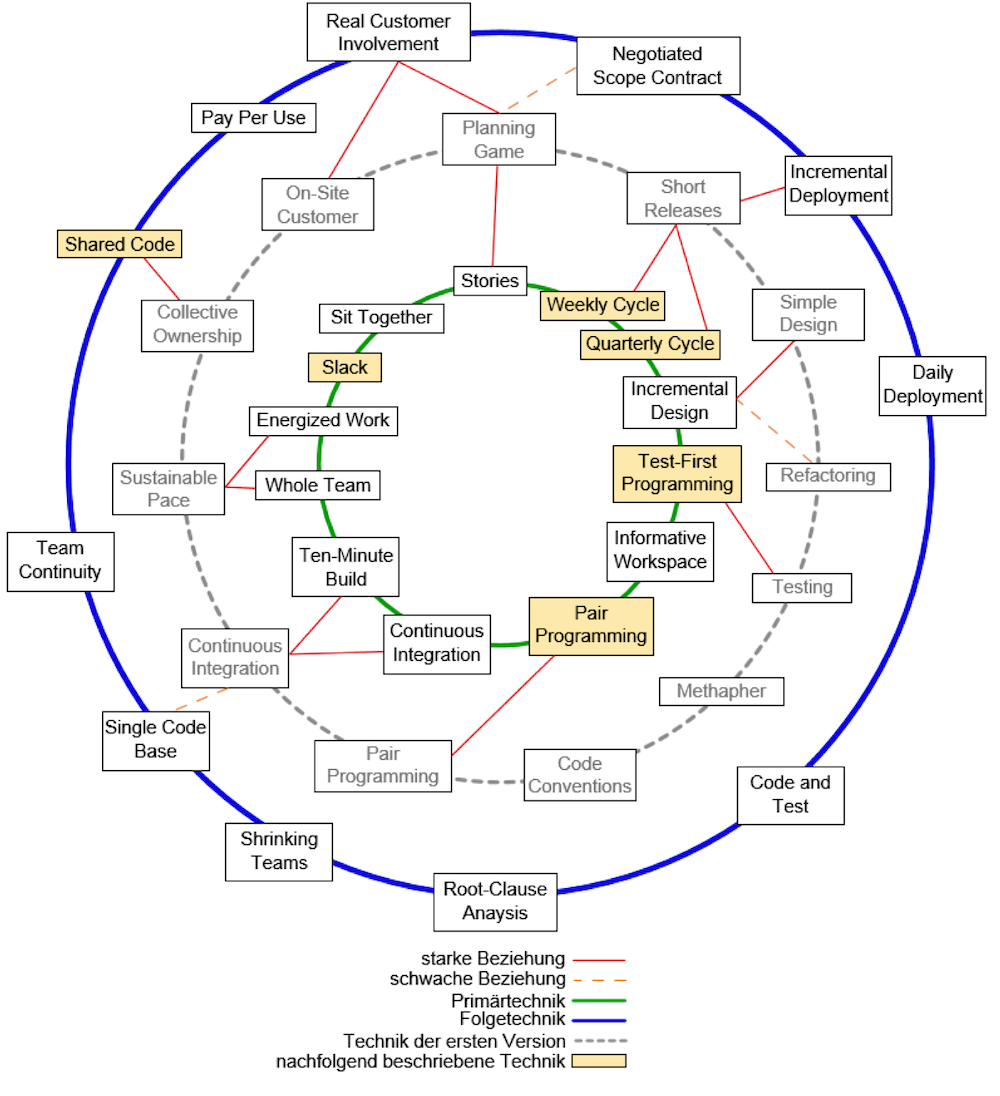}
	\caption{XP Techniques (from \cite{franz2012extreme})}
	\label{fig:XPTech}
\end{figure}

Extreme Programming \cite{beck2000extreme}, familiarly known as XP, originated in 1999 and is a style of software development that focuses on excellent application of programming techniques, clear communication, and teamwork to produce quality software at a sustainable pace. While XP began as a methodology addressing small teams working on internal projects, teams worldwide have used XP for shrink-wrap, embedded, and large-scale projects as well. XP additionally adapts to vague or rapidly changing requirements, but can also be successfully used in those cases in which requirements do not appear to be volatile, such as porting projects. Figure \ref{fig:XPTech} provides a visual summary of the different techniques used in Extreme Programming.

XP is based on the following core values\footnote{\url{http://www.extremeprogramming.org}}: 

\begin{itemize}

\item \textit{Simplicity}: we will do what is needed and asked for, but no more.
\item \textit{Communication}: everyone is part of the team and we communicate face to face daily.
\item \textit{Feedback}: we will take every iteration commitment seriously by delivering working software.
\item \textit{Respect}: everyone will give and feel the respect they deserve as a valued team member.
\item \textit{Courage}: we will tell the truth about progress and estimates.
\end{itemize}

\subsubsection{Kanban}
\label{sec:Kanban}

\begin{figure}[h]
	\includegraphics[width=\columnwidth]{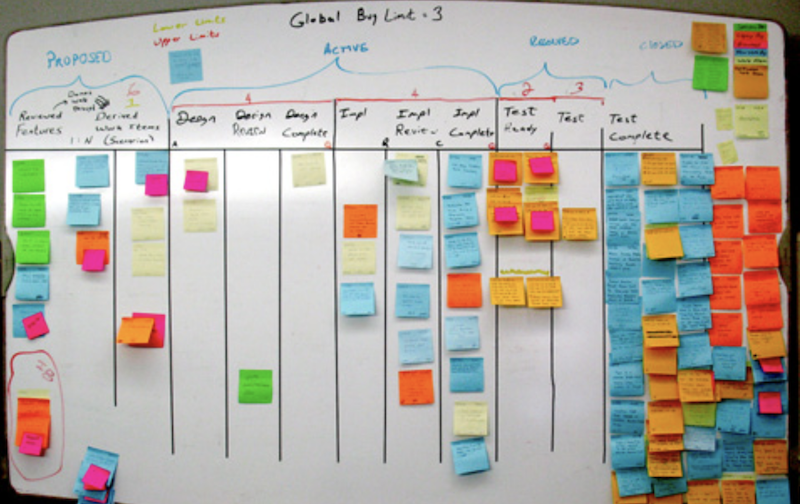}
	\caption{Example of Kanban (from \cite{anderson2010kanban})}
	\label{fig:KanbanExample}
\end{figure}

In software development, Kanban is an agile methodology that places emphasis on continual delivery without overloading the team by limiting work in progress. It is based on a system originally developed in 1980 by the industrial engineer Taiichi Ohne at Toyota to improve manufacturing efficiency \cite{ohno1988toyota,anderson2010kanban}. This system consists of a few kanban (or cards) equivalent to the (agreed) capacity of a system, as shown in the example in Figure \ref{fig:KanbanExample}. One card is attached to one piece of work. Each card acts as a signaling mechanism. A new piece of work can be started only when a card is available. This free card is attached to a piece of work and follows it as it flows through the system. When there are no more free cards, no additional work can be started. Any new work must wait in a queue until a card becomes available. When some work is completed, its card is detached and recycled. With a card now free, a new piece of work in the queuing can be started. 

In software development, each card represents a work item. These cards can be virtual or physic. Card walls have become a popular visual control mechanism in Agile software development. 

Kanban emerged as a methodology in software development at Corbis from 2006 to 2008 and has continued to evolve in the wider Lean software development community in the years since \cite{anderson2010kanban}. Today, it is used to limit a team's work-in-progress to a set capacity and to balance the demand on the team against the throughput of their delivered work. This system has been shown to improve customer satisfaction through regular, dependable, high-quality releases of valuable software. It has also been shown to improve productivity, quality, and lead times.

\begin{figure*}[!t]
	\centering
	\includegraphics[width=0.8\textwidth]{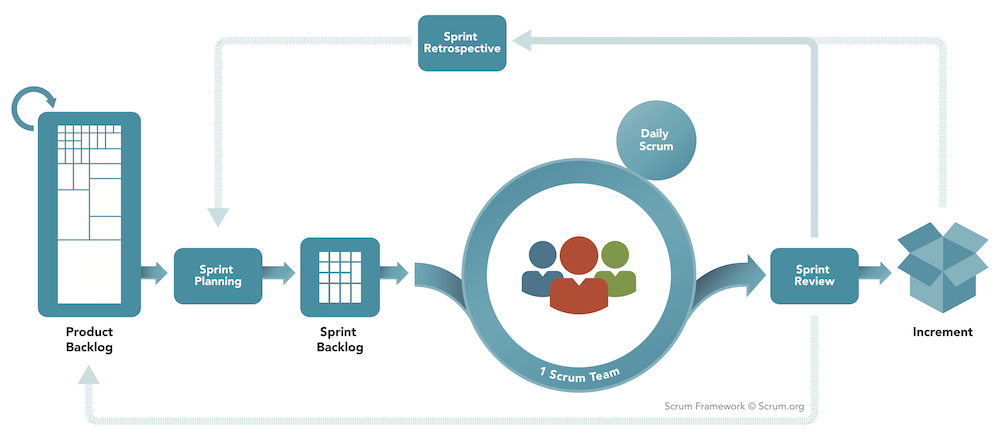}
	\caption{Scrum framework (from https://www.scrum.org)}
	\label{fig:ScrumModel}
\end{figure*}

\subsubsection{Scrum}
\label{sec:Scrum}

Scrum is an agile iterative and incremental product delivery method that uses frequent feedback and collaborative decision making \cite{sliger2011agile}. It is based on a 1986 paper written by Hirotaka Takeuchi and Ikujiro Nonaka for the Harvard Business Review titled ``The New New Product Development Game'' \cite{takeuchi199816}. In this paper, the authors used the sport rugby as a metaphor to describe the benefits of self-organizing teams in innovative product development and delivery. Jeff Sutherland, Ken Schwaber, and Mike Beedle took the ideas from this paper and applied it to their field of software development. They called their new method Scrum, after the rugby term that describes how teams form a circle and go for the ball to get it back into play again. They first applied this method at Easel Corporation in 1993. Schwaber and Beedle wrote about their experiences in their book \textit{Agile Software Development with Scrum} in 2002 \cite{schwaber2002agile}, which was followed by Schwaber's book \textit{Agile Project Management with Scrum} in 2004 \cite{schwaber2004agile}. 

The core element of Scrum is a Sprint, a time-box of one month or less during which a ``Done'', useable and potentially releasable product Increment is created. A new Sprint starts immediately after the conclusion of the previous Sprint \cite{schwaber2016scrum}. Figure \ref{fig:ScrumModel} shows the events, roles and artifacts of the scrum framework.

Scrum prescribes 4 formal events: 

\begin{enumerate}
\item \textit{Sprint Planning}: an event with a maximum duration of eight hours in which a one-month Sprint is planned.
\item \textit{Daily Scrums}: a 15-minute time-boxed event that allows the Development Team to synchronize activities. 
\item \textit{Sprint Review}: a four-hour revision meeting held at the end of the Sprint to inspect the Increment.
\item \textit{Sprint Retrospective}: a three-hour meeting that occurs after the Sprint Review and prior to the next Sprint Planning.
\end{enumerate}

Scrum defines three types of roles:
\begin{enumerate}
\item The \textit{Team} consists of professionals who do the work of delivering a potentially releasable Increment of ``Done'' product at the end of each Sprint.
\item The \textit{ScrumMaster} is responsible for ensuring that Scrum is understood and enacted.
\item The \textit{Product Owner} is responsible for maximizing the value of the product and the work of the Development Team.
\end{enumerate}

Scrum contains 3 types of artifacts that represent work or value to provide transparency and opportunities for inspection and adaptation:

\begin{enumerate}
\item \textit{Product Backlog}: an ordered list of everything that might be needed in the product.
\item \textit{Sprint Backlog}: the set of Product Backlog items selected for the Sprint.
\item \textit{Increment}: the sum of all the Product Backlog items completed during a Sprint.
\end{enumerate}

\begin{figure*}[!t]
	\centering
	\includegraphics[width=0.8\textwidth]{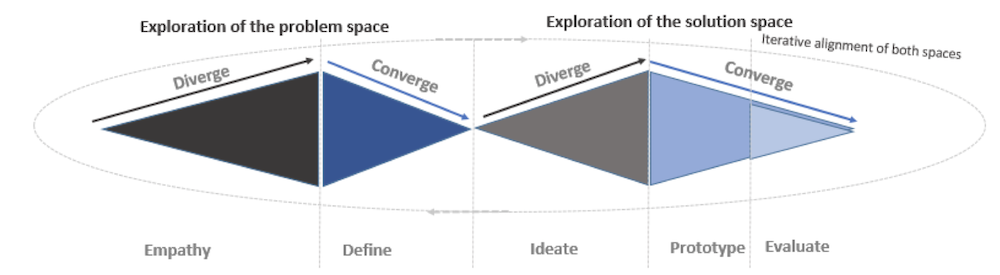}
	\caption{Design Thinking Process (adapted from \cite{faste1993integrating})}
	\label{fig:DesignThinkProc}
\end{figure*}

\subsection{Creativity}
\label{sec:Creativity}

The concept of creativity has been studied from many points of views for decades, and has its roots in the 1950s \cite{runco2012standard}. Creativity, as first mentioned by Barron \cite{barron1955disposition} and Stein \cite{stein1953creativity} requires both originality and effectiveness. Sternberg and Lubart \cite{sternberg1999concept} define creativity as ``the ability to produce work that is both novel (i.e. original, unexpected) and appropriate (i.e. useful, adaptive concerning task constraints)''.

M. Boden \cite{boden2004creative} defines creativity as ``the ability to come up with ideas or artefacts that are new, surprising and valuable''.  According to her research, there are three types of creativity: combinational, exploratory and transformational. Combinational creativity consists in generating unusual (and interesting) combinations of common ideas, while exploratory creativity consists in exploring the different possibilities that exist within a field of knowledge. Finally, transformational creativity completely redefines a field of knowledge or produces a change in the dominant paradigms.

In contrast with the idea that creativity belongs exclusively to a selected few with this ability, there are other theories, such as that proposed by Kaufman and Beghetto \cite{kaufman2009beyond}, which have identified four levels of creativity: Big-C, that consists of clear-cut, eminent creative contributions found in works of creative genius; Little-c or everyday creative activities for non-experts; Mini-c, new and personally meaningful interpretations of experiences, actions and events; and Pro-c, found in professional level expertise in non-creative fields. This theory shows us that almost anyone with the right tools, experience or under certain circumstances could provide creative contributions. 

It is worth noting that none of the primary studies considered in this review explicitly consider this distinction between levels of creativity. However, most of them are focused on leveraging creativity in the agile software development in professional environments. We can conclude then that they deal mainly with Pro-c creativity (found in professional level expertise in non-creative fields).

Within the field of Design, creativity has been widely considered as an essential prerequisite to bring about innovative ideas \cite{snider2011studying}, and some popular approaches, such as Design Thinking, have emphasized how the role of design in development has changed from having a mere focus on the aesthetic aspects and attractiveness of products to delivering innovative ideas in order to create those products \cite{lindberg2011design,brown2008design,brown2009change,martin2009design}. A brief introduction to Design Thinking will be provided in Section \ref{sec:DesignThink}. 

With regard to Requirements Elicitation, during the last decade, engineers have intensively started to underline the importance of creative thinking in requirements elicitation as a decisive factor for the construction of competitive and innovative products \cite{maiden2001requirements,maiden2010requirements}. We shall provide details on this in Section \ref{sec:CreatReqEng}.

\subsubsection{Creativity in Design - Design Thinking}
\label{sec:DesignThink}

Design as a way of thinking is an innovative approach that started in the late 60s \cite{simon1996sciences} and expanded in the 80s and 90s \cite{faste1993integrating,rowe1987g}. Unlike analytical thinking, which is based on data sources and past knowledge to predict future needs, design thinking is proposed as a creative and non-lineal problem-solving approach. 

The main foundations of design thinking are a focus on human behavior, a lot of prototyping and testing, iteration, flexibility, collaboration and multidisciplinary teams \cite{plattner2010introduction}.

Above all, Design Thinking is a human centered approach focused on the users and their needs, which aims to ensure that the solution developed meets a real user need. Since this approach deals with both the problem and the solution as something to be explored, the process requires diverging on many possible solutions (Empathy and Ideate phases) and converging on a focused direction (Define, Prototype and Evaluate phases) \cite{lindberg2011design,plattner2010introduction,clark2008unleashing}. Figure \ref{fig:DesignThinkProc} shows the Design Thinking process.

The first phase, ``Empathy'', consists in a set of activities aimed at understanding users' needs by observing use cases or scenarios. Immediately after the environmental and the physical and emotional needs of the users have been understood, the information that has been collected is analyzed and summarized using tools such as personas and empathy maps. This phase of the ``Define'' process is responsible for bringing clarity and focus to the design space. The next stage, ``Ideate'', focuses on the generation of new ideas in order to avoid obvious solutions and thus increase the potential for innovation. The next phase, ``Prototype'', consists in the iterative generation of artifacts that support the elaboration and evaluation of product concepts with the goal of discovering which ideas proposed during the ``Ideate'' phase are right or wrong. These prototypes are tested by the users during the ``Evaluate'' phase, which aims to show and learn from the users in order to develop better prototypes.

\subsubsection{Creativity in Requirements Engineering}
\label{sec:CreatReqEng}

According to Maiden et. al \cite{maiden2010requirements}, creativity in Requirements Engineering is the capture of requirements that are new to the project stakeholders but may not be historically new to humankind. A number of works have been published in the field of creativity in Requirements Engineering during the last decade. Bhowmik et al. \cite{bhowmik2014automated} propose a framework that extracts familiar ideas from the requirements using topic modeling and creates novel and innovative requirements by obtaining unfamiliar idea combinations. Nguyen and Shanks \cite{nguyen2009framework} have focused on the theoretical side by presenting theoretical fra\-me\-works. Murukannaiah et al. \cite{murukannaiah2016acquiring} propose a sequential Crowd Requirements Engineering process by exploiting human factors, where workers from one stage review requirements from the previous stage and produce additional requirements. Conboy et al. \cite{conboy2009creativity} present a creativity theory in agile methods. In addition, Saha et al. \cite{saha2012systematic} have focused on identifying which creativity techniques are important in requirements elicitation. Svenson et al. \cite{svensson2015creativity} have studied the empirical evidence that exists as regards the use of creativity techniques during the requirements elicitation process and Horkoff et al. \cite{horkoff2015creativity} combine creativity techniques with the structure of goal models, using the contents of the model to facilitate creative thought, in order to enhance software Requirements Engineering.

Furthermore, Lemos et al. \cite{lemos2012systematic} carried out a mapping study on creativity in Requirements Engineering that confirms the relevance of treating requirements as a creative problem-solving exercise. This study comes to the conclusion that authors are proposing new solutions to promote the use of creativity in the requirements elicitation process, in addition to evaluating these proposals in industrial case studies and experiments. The study also discovers that the majority of studies focus on the requirement elicitation phase. It was for this reason that we decided to focus on requirement elicitation rather than on requirements engineering in general.

Although we have found works that analyze agility within Requirements Engineering, such as that presented by Schon et al. \cite{schon2017agile} or creativity in requirements engineering of Lemos et al. \cite{lemos2012systematic} we have not found a systematic review that makes an approach combining creativity in requirements elicitation that specifically looks into agile software development. 

\section{Systematic literature review process}
\label{sec:SLRProcess}
This section describes the process followed in order to carry out the systematic review presented in this paper. It explains the methodology, the search strategy and the procedures that have been carried out to select and extract information.

\subsection{Methodology of the Systematic Literature Review}
\label{sec:MethSLR}

A systematic literature review \cite{kitchenham2004procedures,biolchini2005systematic} is a means of identifying, evaluating and interpreting all available research that is relevant to a particular research question, topic area, or phenomenon of interest in a methodic and reproducible manner. Because of the large number of available sources and the hundreds of thousands of pieces of existing data, a systematic review must be carried out by following a rigorous method. Barbara Kitchenham proposed a method with which to perform a systematic literature review in software engineering that consists of a three-step process: planning, execution and result analysis \cite{kitchenham2004procedures}.

\begin{figure}[h]
	\includegraphics[width=\columnwidth]{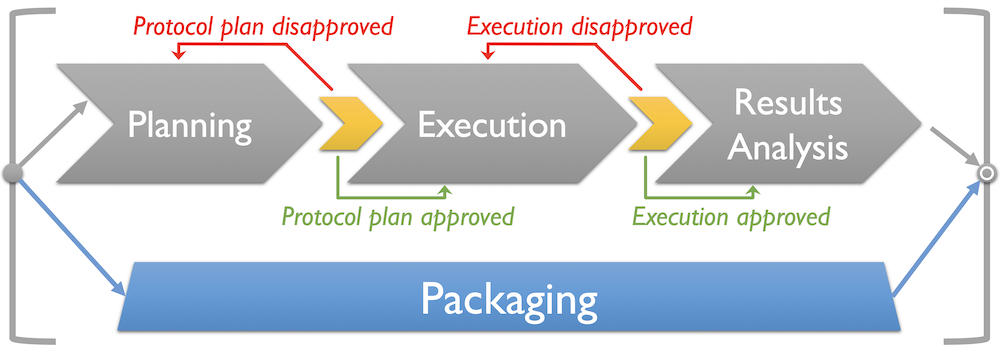}
	\caption{Systematic Review Process (adapted from \cite{biolchini2005systematic})}
	\label{fig:SRProcess}
\end{figure}

However, this method is described at a relatively high level without considering the impact of question type on the review procedures, or providing a detailed specification of the mechanisms that are needed to undertake metaanalysis. It was for these reasons that Biolchini et al. \cite{biolchini2005systematic} proposed a new process based on Kitchenham's proposal to perform a systematic review in which they presented a new approximation composed of four stages: planning, execution, result analysis and packaging. This process is described in Figure \ref{fig:SRProcess}.

The research objectives are listed and a review protocol was defined during the planning phase. This protocol specifies the central research question and the methods that will be used to execute the review. The execution stage involves the identification of primary studies, along with selection and evaluation in accordance with the inclusion and exclusion criteria established in the review protocol. During the result analysis phase, data from the primary studies is extracted and summarized. Concurrently, during these phases, all of the obtained results are stored. This storage is known as the packaging phase, which is performed throughout the whole process. Additionally, there are two checkpoints (represented by diamonds) that allow the detection of problems or errors after the planning and execution phases have been completed \cite{biolchini2005systematic}.

In order to follow this new proposal, Biolchini et al. provided Software Engineering researchers with a template that would serve as a guideline when conducting systematic reviews. Since this template is focused on systematic reviews based on empirical studies, and the objective of this paper is to carry out a literature analysis of a very new area in which proper empirical evidence might be very limited, we have followed this template but have excluded the following subsections: control, population, experimental design, results statistical calculus, sensitivity analysis and plotting.

\subsection{Objective of the systematic review and definition of research questions}
\label{sec:ObjectiveSR}

The aim of the present study is to employ a systematic review to investigate the state of the art of the techniques or methodologies that apply creative approaches to requirements elicitation in Agile Software Development, along with the benefits and limitations and the strength of evidence of these approaches.

Agile Methodologies have already proven beneficial as regards reducing the delivery time of working software, improving customer collaboration, estimating time and handling defects processes \cite{dybaa2008empirical}. Some studies, however, suggest that the existing agile approaches, when used during the requirements analysis phase, are prone to focus only on scoping and simplicity rather than on problem solving and discovery \cite{turk2014limitations,hosbond2008misfit}. Considering that requirements elicitation plays a determining role in the development process and that it is characterized by ambiguity, uncertainty, and unpredictability \cite{schon2017agile}, it has been proposed that agile processes be extended with creative techniques in order to manage complex software developments \cite{hollis2013extending,maiden2010requirements,mahaux2013collaborative,lemos2012systematic}. We have attempted to identify the state of the art in this topic by defining the following research questions (RQ):

\begin{itemize}
\item \textit{RQ1}. What methodologies or techniques that involve creative approaches in requirements elicitation (RE) in agile software development (ASD) exist? 
\item \textit{RQ2}. What empirical evidence is there of the application of creative techniques for RE in ASD?
\item \textit{RQ3}. What is known about the benefits and limitations of using creative approaches for RE in ASD? 
\item \textit{RQ4}. In what sort of organizations or projects using ASD could the use of these creative approaches be most suitable? 
\item \textit{RQ5}. Which agile methodologies are being used to integrate creative techniques into agile software development RE processes?

\end{itemize}

\subsection{Data sources and query strings}
\label{sec:DataSourcesQS}

Once the research questions had been established, a set of keywords that matched the research objectives was selected. This set of keywords covered three main domains: ``Creativity'', ``Requirements'' and ``Agile''. Alternative spellings and synonyms were identified for each domain. The complete set of keywords can be seen in Table \ref{tab:primaryStudies}. The selected keywords were subsequently connected with Boolean operators to create a search query as follows:
(``design thinking'' OR creative OR creativity) AND (requirements OR ``product backlog'' OR preconditions OR specifications OR modeling OR analysis OR ``user stories'') AND (agile OR scrum OR Kanban OR lean OR crystal OR xp OR ``extreme programming'').

The objective was to create a search query that would cover at least one term of each domain. This search query was then adapted to the syntax of the different search engines.

\begin{table}[h]
	\caption{Set of keywords}
	\label{tab:primaryStudies}
	\centering
	\begin{tabular}{ccc}
		\toprule
		\textbf{Requirements} & \textbf{Creativity}   &   \textbf{Agile}        \\ \midrule
		 Requirement & Creativity & Agile \\
		 Product Backlog & Creative & Scrum   \\
		 Precondition & & Kanban \\
		 Specification & & Lean \\
		 Modeling & & Crystal \\
		 Analysis & & XP\\
		 User Stories & & Extreme Programming\\
		\bottomrule
	\end{tabular}
\end{table}

A fourth block of keywords was initially included in the query related to ``Information Systems and Software Engineering''. However, as evidenced by the search results, it was later observed that this block was overcomplicating the search query without providing additional relevant results. Something similar occurred while testing the extra search query terms, such as other types of agile methodologies (feature-driven development, Dynamic systems development method) or synonyms such as ``agility''. Having said that, owing to the limitation of terms in some search engines or the inability to attain additional relevant results with those synonyms, we decided to keep the query as simple as possible.

During this stage, data sources were identified and a manual search process was executed in specific electronic databases. The results obtained were analyzed and the data sources were sorted. We initially considered a set of digital libraries which was later considerably reduced, because after obtaining the results some digital libraries did not provide us with relevant information. The final list of data sources that has been employed for this systematic review is shown in Table \ref{tab:selectedSearchSources}.

\begin{table}[h]
	\caption{Set of selected data sources}
	\label{tab:selectedSearchSources}
	\centering
	\begin{tabular}{@{}ll@{}}
		\toprule
		\textbf{Source}       & \textbf{Website}            \\ \midrule
		ACM & http://portal.acm.org \\
		Google Scholar & http://scholar.google.com \\
		IEEE Xplore & http://ieeexplore.ieee.org \\
		ISI Web of Science & http://www.isiknowledge.com/ \\
		Science Direct & http://www.sciencedirect.com \\
		Scopus & https://www.scopus.com \\ \bottomrule 
		
	\end{tabular}
\end{table}

\subsection{Study selection}
\label{sec:Study selection}

\begin{figure*}[!t]
	\centering
	\includegraphics[width=\textwidth]{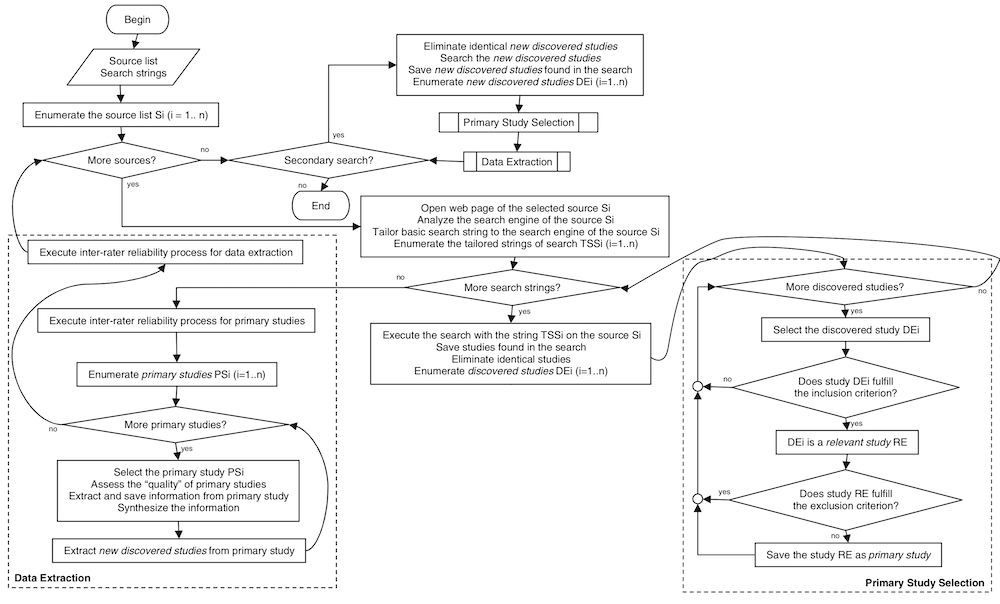}
	\caption{Procedure for execution os the systematic review (from \cite{pino2008software})}
	\label{fig:SLRProcedure}
\end{figure*}

In order to proceed with the primary study selection, this systematic review followed the procedure proposed by Pino, Garc\'ia and Piattini \cite{pino2008software}, which is based on the proposal of Biolchini et al. \cite{biolchini2005systematic}. This procedure is described in Figure \ref{fig:SLRProcedure}.

\begin{figure}[h]
	\advance\leftskip 1.5cm
	\includegraphics[width=0.6\columnwidth]{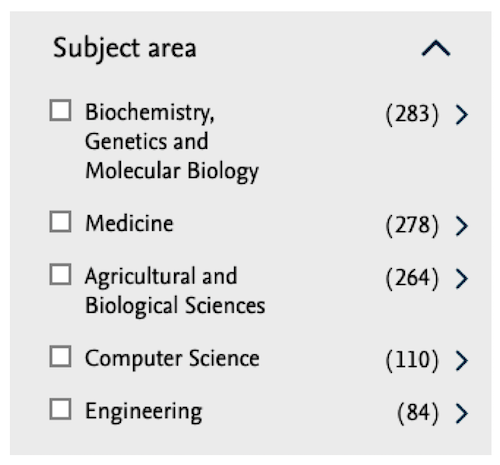}
	\caption{Example of Filter Option in Scopus}
	\label{fig:ScopusFilter}
\end{figure}

The search query described in the previous section was adapted to each of the selected sources by considering the following criteria:

\begin{enumerate}
	
\item Restriction by Title \& Abstract: if the data source allowed us to restrict the search query only to the title and abstract or title, abstract and keywords, the query was modified accordingly. For example, in the ACM Digital Library we added the prefix \textit{acmdlTitle}: to search for a keyword only within the title of a study and the prefix \textit{recordAbstract}: to search only in the abstract section. Similarly, in ScienceDirect and Scopus we included the prefix \textit{TITLE-ABS-KEY} before each keyword. This is an example of an adapted search query:\\

	\fbox{\begin{minipage}{22em}
			\centering
			\scriptsize{( TITLE-ABS-KEY (``design thinking'') OR TITLE-ABS-KEY ( creative) OR TITLE-ABS-KEY ( creativity ) ) AND ( TITLE-ABS-KEY ( requirements ) OR TITLE-ABS-KEY (``product backlog'') OR TITLE-ABS-KEY ( preconditions) OR TITLE-ABS-KEY ( specifications) OR TITLE-ABS-KEY ( modeling ) OR TITLE-ABS-KEY ( analysis ) OR TITLE-ABS-KEY (``user stories'')) AND ( TITLE-ABS-KEY ( agile ) OR TITLE-ABS-KEY ( scrum ) OR TITLE-ABS-KEY ( kanban ) OR TITLE-ABS-KEY ( lean ) OR TITLE-ABS-KEY ( crystal ) OR TITLE-ABS-KEY ( xp ) OR TITLE-ABS-KEY ( ``extreme programming'') )}
	\end{minipage}}\\
	\label{box:queryString}

	\item Restriction by field of study: if the data source offered an option to filter results to only computer science related studies, the initial list of results was filtered (see Figure \ref{fig:ScopusFilter}).
\end{enumerate}

The search strings were run in the selected sources. The execution of these searches provided a total of 1451 results. Table \ref{tab:searchResults} shows the results obtained after executing the adapted search queries.

\begin{table}[h]
	\caption{Search results overview}
	\label{tab:searchResults}
	\centering
	\begin{tabular}{lcc}
		\toprule
		\textbf{Search space} & \textbf{Search results}   &   \textbf{Filtered results}        \\ \midrule
			ACM & 110 & 110 \\
			Google Scholar & 937 & 937 \\
			IEEE Xplore & 61 & 61 \\
			ISI Web of Science & 1259 & 223 \\
			Science Direct & 24 & 10 \\
			Scopus & 554 & 110 \\
			\textbf{All Libraries} & \textbf{2945} & \textbf{1451} \\
		\bottomrule
	\end{tabular}
\end{table}

An initial set of relevant studies was selected by reading the title and abstract of all the studies obtained from the web search engines, and these were evaluated according to the inclusion and exclusion criteria. The studies were eligible for inclusion if they fulfilled the following criteria: all kinds of studies related to the research topic will be selected if, after analyzing title and abstract of the document, it is obvious that the study presents some kind of creative approach or technique used for gathering requirements in agile software development. All documents that did not fulfill these criteria were discarded.

It was found that, of the 1451 gathered studies, only 51 were relevant, i.e., 4 per cent. The number of relevant studies that were found in each digital library is presented in the third column of Table \ref{tab:searchResultsDetailed}. These numbers are detailed as percentages in the 4\textsuperscript{th} and 5\textsuperscript{th} columns.

\begin{table}[h]
	\caption{Details of search results}
	\label{tab:searchResultsDetailed}
	\centering
	\begin{tabular}{lcccc}
		\toprule
		\textbf{Source} & \textbf{Results}   &   \textbf{Relevant} & \textbf{\% of rel.} & \textbf{\% of all rel.} \\ \midrule
		{\scriptsize ACM} & 110 & 7 & 6\% & 14\% \\
		{\scriptsize Google Scholar} & 937 & 17 & 2\% & 33\% \\
		{\scriptsize IEEE Xplore} & 61 & 6 & 10\% & 12\% \\
		{\scriptsize ISI Web} & 223 & 9 & 4\% & 18\% \\
		{\scriptsize Science Direct} & 10 & 0 & 0\% & 0\% \\
		{\scriptsize Scopus} & 110 & 12 & 11\% & 24\% \\
		\textbf{All libraries} & \textbf{1451} & \textbf{51} & \textbf{4\%} & \textbf{100\%} \\
		\bottomrule
	\end{tabular}
\end{table}

There is a noteworthy difference between the number of obtained results and the number of relevant results. It was observed that even though many of the studies that were acquired during the search contained some words from the search query, their scope was not related to the research questions presented in this systematic literature review (Section \ref{sec:ObjectiveSR}) and they were, therefore, excluded. In addition, it is worth mentioning that many of the collected works were related to creative approaches but not within the specific context of agile development or to requirements analysis in particular.

It is important to consider that one specific study may be found in several search spaces at the same time. This is denominated as a relevant duplicated study and should be filtered. Once the duplicated studies had been removed, 17 relevant studies remained (see Table \ref{tab:overviewRelevantStuies}). None of these studies fulfilled the exclusion criteria and they were, therefore, all selected as primary studies. Appendix I contains the full list of the primary studies.

\begin{table}[h]
	\caption{Overview of relevant studies}
	\label{tab:overviewRelevantStuies}
	\centering
	\begin{tabular}{lcc}
		\toprule
		\textbf{Relevant studies} & \textbf{\# of studies}   &   \textbf{Percentage}        \\ \midrule
		Relevant studies & 51 & 100\% \\
		Duplicated relevant studies & 34 & 67\% \\
		\textbf{Non-duplicated relevant studies} & \textbf{17} & \textbf{33\%} \\
		\bottomrule
	\end{tabular}
\end{table}

\subsection{Studies quality assessment planning}
\label{sec:StudiesQuality}

In order to avoid biases and ensure the relevance of the selected studies, we undertook a quality assessment. Since standard quality criteria that are valid for all systematic reviews do not exist, we compiled questions from the suggestions proposed by Kitchenham in \cite{kitchenham2008personal}. The proposed checklist in this systematic review assesses the individual studies in terms of rigor, credibility and relevance (see Table \ref{tab:QualityAR}, \ref{tab:QualityCred} and \ref{tab:QualityRelev}).

\begin{table*}[htb]
	\centering
	\caption{Quality assessment - Rigor}
	\label{tab:QualityAR}
	\resizebox{\textwidth}{!}{
		\begin{tabular}{lllcl}
			\toprule
			\multicolumn{1}{c}{\textbf{Category}} & \multicolumn{1}{c}{\textbf{Item}} & \multicolumn{1}{c}{\textbf{Assessment criteria}} & \multicolumn{1}{c}{\textbf{Score}} & \multicolumn{1}{c}{\textbf{Description}} \\ \midrule
			
			\multirow{9}{*}{\begin{tabular}[c]{@{}l@{}}Rigor:\\ Has a thorough and\\ appropriate approach\\ been applied to key\\ research methods\\ in the study?\end{tabular}} & \multirow{3}{*}{QA1} & \multirow{3}{*}{\begin{tabular}[c]{@{}l@{}}Does the paper include\\ a clear statement of the\\ aims of the study? \end{tabular}} & 2 & Yes, aims of the study are clearly described \\ \cline{4-5}
			
			& & & 1 & Partially, aims are described but unclearly \\ \cline{4-5}
			
			& & & 0 & No, aims are not described \\ \cline{2-5}
			
			& \multirow{3}{*}{QA2} & \multirow{3}{*}{\begin{tabular}[c]{@{}l@{}}Does the study present\\ a detailed description of\\ the approach? \end{tabular}} & 2 & \begin{tabular}[c]{@{}l@{}}Yes, the approach is well described and can be applied \end{tabular} \\ \cline{4-5}
			
			& & & 1 & \begin{tabular}[c]{@{}l@{}}Partially, the approach is difficult to understand and to replicate \end{tabular} \\ \cline{4-5}
			
			& & & 0 & No, details are missing \\ \cline{2-5}
			
			& \multirow{3}{*}{QA3} & \multirow{3}{*}{\begin{tabular}[c]{@{}l@{}}Does the study follow a syste-\\ matic methodology that can\\ be applied in another setting?\end{tabular}} & 2 & Yes, if follows a systematic methodology \\ \cline{4-5}
			
			& & & 1 & Partially, if follows a methodology but it is not fully described \\ \cline{4-5}
			
			& & & 0 & No, it does not follow a systematic methodology \\
		\bottomrule
	\end{tabular}%
	}
\end{table*}

\begin{table*}[htb]
	\centering
	\caption{Quality assessment - Credibility}
	\label{tab:QualityCred}
	\resizebox{\textwidth}{!}{
		\begin{tabular}{lllcl}
			\toprule
			\multicolumn{1}{c}{\textbf{Category}} & \multicolumn{1}{c}{\textbf{Item}} & \multicolumn{1}{c}{\textbf{Assessment criteria}} & \multicolumn{1}{c}{\textbf{Score}} & \multicolumn{1}{c}{\textbf{Description}} \\ \midrule
			
			\multirow{12}{*}{\begin{tabular}[c]{@{}l@{}}Credibility:\\ Are the findings\\ well-presented and\\ meaningful? \end{tabular}} &

			\multirow{3}{*}{QA4} & \multirow{3}{*}{\begin{tabular}[c]{@{}l@{}}Is the reporting clear and coherent? \end{tabular}} 
			
			& 2 & Yes, the findings are clearly described \\ \cline{4-5}
			
			& & & 1 & Partially, the findings are described but are not easy to understand \\ \cline{4-5}
			
			& & & 0 & No, the findings are difficult to understand \\ \cline{2-5}

			& \multirow{3}{*}{QA5} & \multirow{3}{*}{\begin{tabular}[c]{@{}l@{}}Is the research process been\\ adequately documented? \end{tabular}} 
			
			& 2 & \begin{tabular}[c]{@{}l@{}}Yes, the research process is fully documented \end{tabular} \\ \cline{4-5}
			
			& & & 1 & \begin{tabular}[c]{@{}l@{}}Partially, some parts of the research process are omitted \end{tabular} \\ \cline{4-5}
			
			& & & 0 & No, the research process is very inadequately documented \\ \cline{2-5}

			& \multirow{3}{*}{QA6} & \multirow{3}{*}{\begin{tabular}[c]{@{}l@{}}Is the proposal validated?\end{tabular}} 
			
			& 2 & Yes, it is validated \\ \cline{4-5}
			
			& & & 1 & Partially, validation is ongoing or data is not significant \\ \cline{4-5}
			
			& & & 0 & No, it is not validated \\ \cline{2-5}

			& \multirow{3}{*}{QA7} & \multirow{3}{*}{\begin{tabular}[c]{@{}l@{}}Are the links between data,\\ interpretation and conclusions\\ clear? \end{tabular}} 
			
			& 2 & \begin{tabular}[c]{@{}l@{}}Yes, results are clearly described \end{tabular} \\ \cline{4-5}
			
			& & & 1 & \begin{tabular}[c]{@{}l@{}}Partially, results are described but not properly linked to data \end{tabular} \\ \cline{4-5}
			
			& & & 0 & No, interpretations and conclusions are unclear or not described at all \\ 
			
			\bottomrule
		\end{tabular}%
	}
\end{table*}

\begin{table*}[htb]
	\centering
	\caption{Quality assessment - Relevance}
	\label{tab:QualityRelev}
	\resizebox{\textwidth}{!}{
		\begin{tabular}{lllcl}
			\toprule
			\multicolumn{1}{c}{\textbf{Category}} & \multicolumn{1}{c}{\textbf{Item}} & \multicolumn{1}{c}{\textbf{Assessment criteria}} & \multicolumn{1}{c}{\textbf{Score}} & \multicolumn{1}{c}{\textbf{Description}} \\ \midrule
			
			\multirow{9}{*}{\begin{tabular}[c]{@{}l@{}}Relevance:\\ How useful are the\\ findings to the\\ software industry\\ and the research\\ community? \end{tabular}} &

			\multirow{3}{*}{QA8} & \multirow{3}{*}{\begin{tabular}[c]{@{}l@{}}Is knowledge or understan-\\ ding been extended by\\ the research? \end{tabular}} 
			
			& 2 & Yes, the research provides new useful knowledge or understanding \\ \cline{4-5}
			
			& & & 1 & Partially, the study expands some knowledge from previous study \\ \cline{4-5}
			
			& & & 0 & No, the research does extend knowledge or understanding \\ \cline{2-5}

			& \multirow{3}{*}{QA9} & \multirow{3}{*}{\begin{tabular}[c]{@{}l@{}}Is the proposal cited by\\ other authors? \end{tabular}} 
			
			& 2 & \begin{tabular}[c]{@{}l@{}}Yes, it is cited by 5 or more authors \end{tabular} \\ \cline{4-5}
			
			& & & 1 & \begin{tabular}[c]{@{}l@{}}Very few, less than 5 articles cited the study \end{tabular} \\ \cline{4-5}
			
			& & & 0 & No, no one cited the study \\ \cline{2-5}

			& \multirow{3}{*}{QA10} & \multirow{3}{*}{\begin{tabular}[c]{@{}l@{}}Does the study present a pro-\\ posal that can be replicated in\\ other organizations/settings? \end{tabular}} 
			
			& 2 & Yes, the proposal can be replicated under different circumstances \\ \cline{4-5}
			
			& & & 1 & Partially, the proposal can be replicated with some restrictions \\ \cline{4-5}
			
			& & & 0 & No, the proposal cannot be replicated \\

			\bottomrule
		\end{tabular}%
	}
\end{table*}

\subsection{Data extraction and synthesis planning}
\label{sec:DataExtractionSynt}

According to the guidelines of Biolchini et al. \cite{biolchini2005systematic}, certain information needs to be extracted from each of the selected studies. To facilitate this task, we employed the Mendeley software to extract basic information, such as the title, authors, publication date, DOI, type of work and abstract. We additionally designated a template for each primary study by adapting the proposal from Pino et al. \cite{pino2008software} to collect the most important characteristics of each study (see Table \ref{tab:DataExtracCharacPS}).

\begin{table}[h]
	\caption{Data extraction - Characteristics of primary studies}
	\label{tab:DataExtracCharacPS}
	\centering
	\begin{tabular}{ll}
		\toprule
		\textbf{Title} & Title of the study \\
		\textbf{Authors} & Authors of the study \\
		\textbf{Published in...} & Journal, magazine or conference \\
		\textbf{Year} & Year of publication \\
		\textbf{Abstract} & Summary of the study \\
		\bottomrule
	\end{tabular}
\end{table}

Besides this basic information, we extracted relevant information that answered each of the research questions (see Table \ref{tab:DataExtracRQ}).

\begin{table}[h]
	\caption{Data extraction - Research questions}
	\label{tab:DataExtracRQ}
	\centering
	\begin{tabular}{ll}
		\toprule
		\textbf{R1} & What Creative Technique? \\
		\textbf{R2} & What Empirical Evidence? \\
		\textbf{R3} & What Benefits and limitations? \\
		\textbf{R4} & Where is, or could be used? \\
		\textbf{R5} & What Agile Methodologies? \\
		\bottomrule
	\end{tabular}
\end{table}

Furthermore, in order to facilitate the synthesis of the studies, we collected information regarding the problems identified by the studies, the objectives, the methodology employed, the conclusions and the future work (see Table \ref{tab:DataExtracFI}).

\begin{table}[h]
	\caption{Data extraction - Further information}
	\label{tab:DataExtracFI}
	\centering
	\begin{tabular}{ll}
		\toprule
		\textbf{Identified Problems} & \begin{tabular}[c]{@{}l@{}}What problems identified by the authors\\ exist in requirements elicitation? \end{tabular} \\ \cline{2-2}
		\textbf{Objective} & \begin{tabular}[c]{@{}l@{}}What are the motivations and aims of the\\ study? \end{tabular} \\ \cline{2-2}
		\textbf{Methodology} & \begin{tabular}[c]{@{}l@{}}What research Methodology is being\\ followed by the study? \end{tabular} \\ \cline{2-2}
		\textbf{Conclusions} & \begin{tabular}[c]{@{}l@{}}What are the conclusions of the study? \end{tabular} \\ \cline{2-2}
		\textbf{Future work} & \begin{tabular}[c]{@{}l@{}}What is the future work suggested by\\ the study? \end{tabular} \\
		\bottomrule
	\end{tabular}
\end{table}

\section{Data extraction}
\label{sec:dataExtraction}
This section summarizes the most important data that has been extracted from the primary studies. A classification of the studies and a quality assessment of them will be presented later in this paper.

\subsection{Overview of studies}
\label{sec:OverviewStudies}

This section summarizes the results of the review of the primary studies.

\subsubsection{Maiden and Hollis}
\label{sec:MaidenHollis}

In [S1] the authors propose to integrate creative thinking into agile processes in order to generate new and useful requirements. The authors first question the effectiveness of agile processes, which very often miss some requirements by just thinking of the simplest solution. Based on previous studies \cite{oza2009building}, they claim that agility is not always sufficient to generate the requirements needed for innovative new software solutions. One of the reasons for this is that the short duration of a sprint may discourage creativity.

\begin{figure*}[h]
	\centering
	\includegraphics[width=0.8\textwidth]{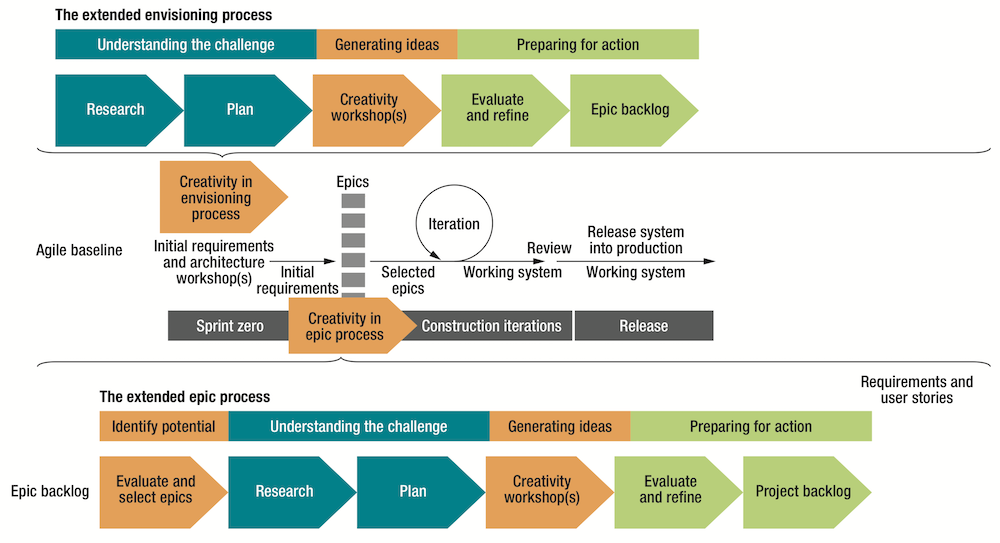}
	\caption{The Extended Envisioning Process [S1]}
	\label{fig:ExtEnvProcess}
\end{figure*}

To solve this problem, they present an extension of the Ambler agile process \cite{ambler2002agile}. This extension consists of two sub processes that introduce creative techniques, as seen in Figure \ref{fig:ExtEnvProcess}. The first creative sub process is included during the envisioning process in sprint zero, between the plan and the evaluation phases. The second creative sub process takes place at the beginning of some sprints, when the epics with the most creative potential are selected and become the focus of the creative activities. During these creative phases, short creativity workshops are run to discover more new ideas. The idea of using creativity workshops to support the creative process in requirements engineering had already been investigated by the same authors in previous studies \cite{maiden2007can,zachos2008inventing,maiden2004integrating,maiden2005integrating,jones2007creativity}.

These creative workshops employ different creative techniques that have been, to a great extent, inspired by or extracted from the works of Higgins and Michalko \cite{higgins1994101,michalko2010thinkertoys}. This work follows a case study research methodology and provides evidence that the method presented could be used in real software projects. However, although several projects in which the technique was applied are mentioned, only one is documented, which consisted in redesigning a television listing website at BBC Worldwide. The results of the experiment suggest that the use of creative techniques to elicit requirements can produce more new requirements. These requirements, however, tend to be seen as less useful, especially in the initial phases. For these reasons, the authors recommend using this type of technique in the search phase of projects rather than to produce incremental change. For future works the authors recommend the use of light\-weight creativity techniques in other agile projects.

\subsubsection{O'Driscoll}
\label{sec:ODriscoll}

\begin{figure}[h]
	\centering
	\includegraphics[width=\columnwidth]{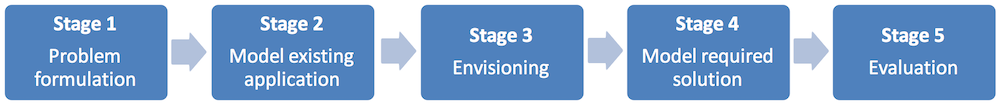}
	\caption{Agile Design Data Modelling process}
	\label{fig:AgileDesign}
\end{figure}

In [S2], the author presents a technique named the Agile Design Data Modelling (ADDAM) process. Motivated by the difficulties of requirements analysis to meet the needs and expectations of end-users, this solution aims to improve the requirements analysis process and the value of the solutions developed by involving agile and design thinking in business oriented data modeling. The process, as shown in Figure \ref{fig:AgileDesign}, consists of five stages: \textit{Problem} \textit{Formulation}, \textit{Model Existing Application}, \textit{Envisioning}, \textit{Model Required Solution}, and \textit{Evaluation}.

\begin{figure}[h]
	\includegraphics[width=\columnwidth]{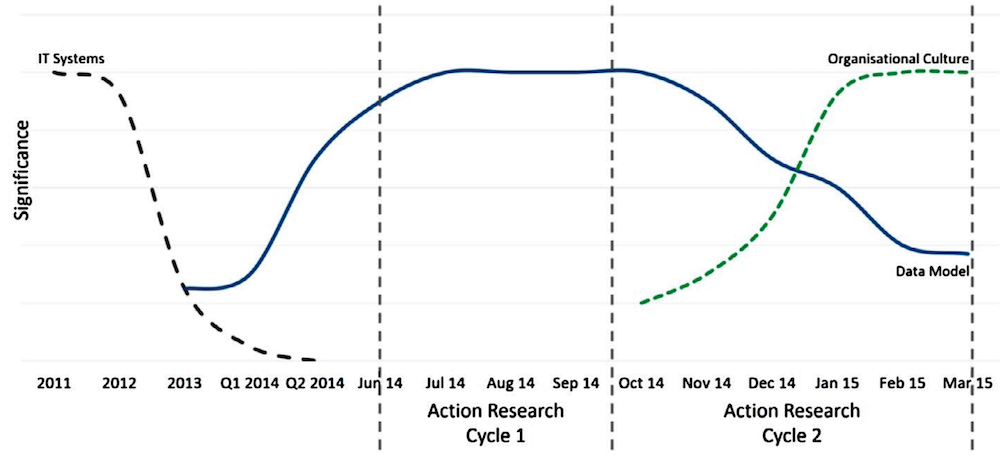}
	\caption{Action research project overview [S2]}
	\label{fig:ActionResearch}
\end{figure}

This work documents an action research study which took place within an organization's IT Project between May 2014 and March 2015 (see Figure \ref{fig:ActionResearch}). No other empirical evidence or studies which apply this technique have been found.

The study suggests that this technique for requirements analysis can be effective at improving business stakeholders' engagement and building a better understanding of the business and its problems. This leads to cost savings and the development of solutions that better meet business needs. It also suggests that the data model presented improves communication between the business and IT groups and also enables faster and more efficient systems development. The study also reports that when it comes to Information Systems development, the non-technical factors around people and organizational culture are much more significant than the technological and data-related factors.

\begin{figure*}[h]
	\centering
	\includegraphics[width=0.8\textwidth]{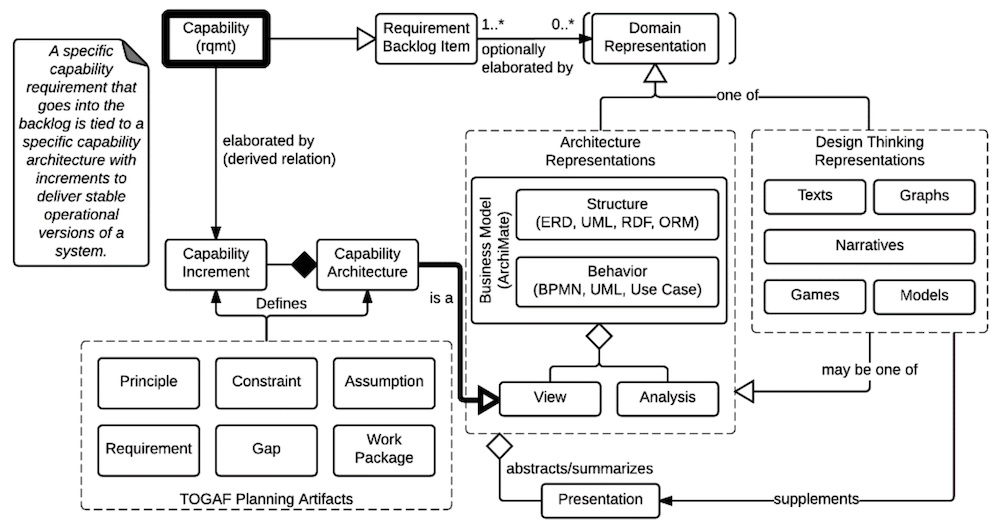}
	\caption{ADDAM Process detailed description [S3]}
	\label{fig:ADDAM}
\end{figure*}

\subsubsection{Gamble}
\label{sec:Gamble}

The main idea behind the technique proposed in [S3] is to align and standardize metamodels in order to strengthen the link between design intent, business and development. The author underlines that in many software development projects, despite recent progress as regards adopting agile methodologies, the output does not always meet the business needs. One of the reasons for this is that each stakeholder uses his or her own set of methods and tools, and these different frameworks are poorly aligned. 

In order to solve this problem, the author examines the different metamodels used by agile development, service design thinking and architecture frameworks and provides a proposal with which to align these metamodels. He illustrates the technique by using examples of the following frameworks: TOGAF (The Open Group Architecture Framework) for enterprise architecture, SAFe (the Scaled Agile Framework) \cite{leffingwell2010agile} for agile enterprise development, and the concepts of service design thinking \cite{stickdorn2011service,lovlie2013service}. The result of linking these frameworks is the model presented in Figure \ref{fig:ADDAM}. This approach offers a design linkage while simultaneously constraining the complexity to a level that is usable by developers, and traceable by architects.

Apart from the examples presented to explain the model, no further evidence is provided by the study. Although there is the potential to tie the development task to the design artifacts from architecting and design thinking activities, there is still very little guidance. The suggested further work is to provide empirical evidence by testing the proposal in different situations of enterprise development.

\subsubsection{Newman and Ferrario}
\label{sec:NewmanFerrario}

The technique proposed in [S4] is based on the integration of design thinking into an agile and participatory framework. It does so by introducing user collaboration from the very beginning, even before requirements are elicited.

\begin{figure}[h]
	\advance\leftskip 1.7cm
	\includegraphics[width=0.55\columnwidth]{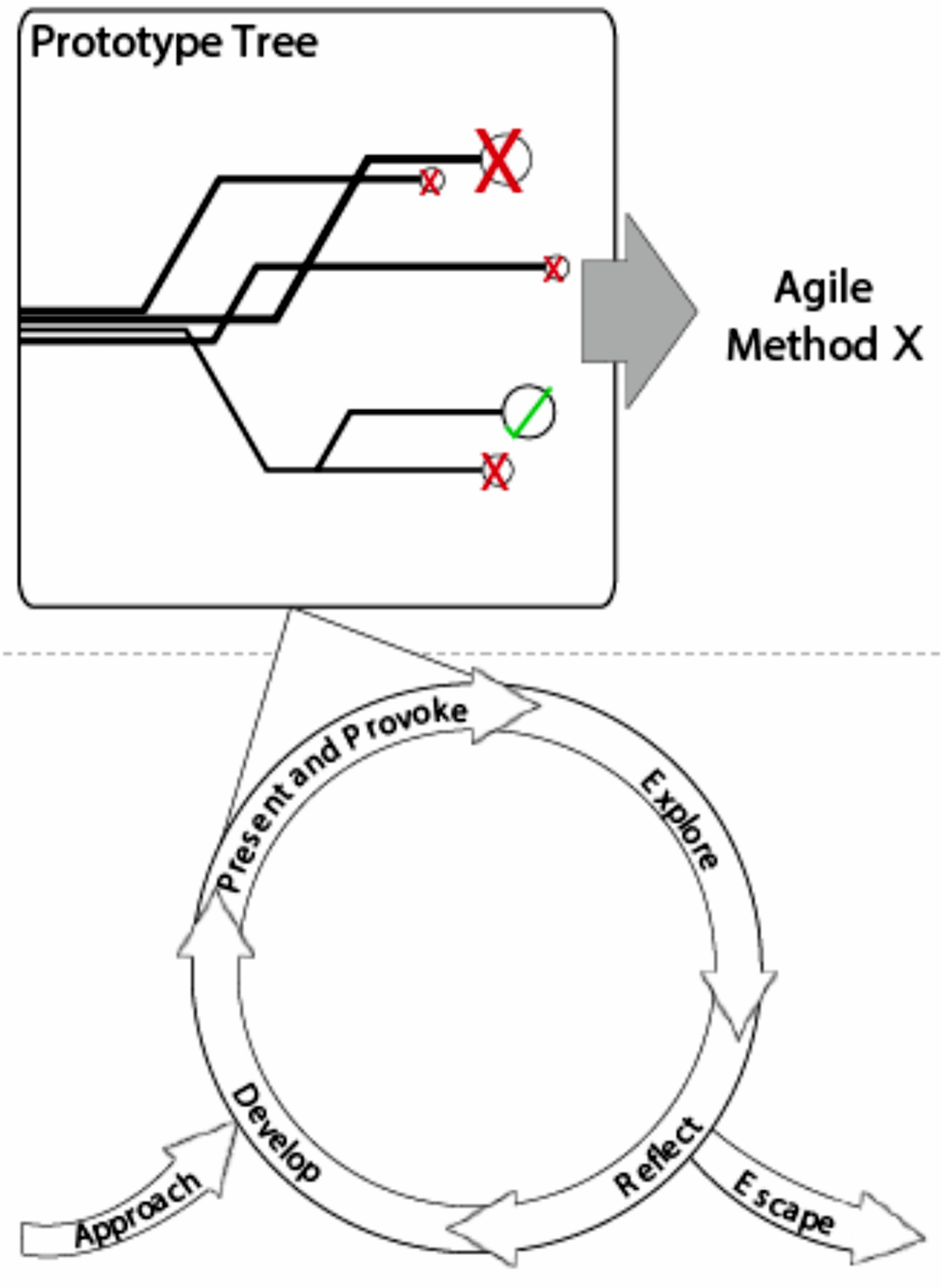}
	\caption{Divingboard process [S4]}
	\label{fig:Divingboard}
\end{figure}

\sloppy
The authors emphasize that complex and undefined social problems cannot be addressed through the use of formal methods and analysis alone. Even though some works have already made use of creativity techniques like creativity workshops to elicit requirements \cite{maiden2004integrating,schlosser2008using}, it has not yet been investigated whether such approaches can be applied to groups with little or no experience in the problem domain. Similarly, the approaches that present artifacts as focal points for participants to use in the design process are not developed on the basis of feedback obtained from the stakeholders. Motivated by these problems, the authors' purpose is to integrate design thinking into an agile and participatory framework that can be applied to complex and not well understood socio-technical problems. This creative approach is called ``Divingboard'' and makes use of creative workshops and physical artifacts to allow participants to better engage with the problem context. An overview of the process can be observed in Figure \ref{fig:Divingboard}.

\sloppy
This work provides empirical evidence of the technique presented in the form of a nine-month case study. This was carried out with a remote Scottish Island community in order to develop a renewable energy forecasting system. The results of the study show that this technique facilitates skill sharing, team building, creative problem exploration, the rapid prototyping of ideas and possible solutions for complex, underspecified and open-ended problems. Nevertheless, it also underlines that users should be open minded and willing to discard and change ideas if the prototype proves unsuccessful.

The study concludes that creativity-driven workshops and physical artifacts are effective tools as regards encouraging the generation of requirements and solutions for complex problems. Furthermore, in this creative problem-solving technique, user participation is key aspect to ensure the success of the agile software development process.

\subsubsection{Patton}
\label{sec:Patton}

The proposal presented in [S5] introduces a technique that consists in using interaction design concepts in an agile development process to help better define requirements. The author identifies that the use of agile methodologies, such as XP, tends to deliver high quality software quickly, but the resulting product does not impress customers or meet their expectations. Very often, the final product ends up offering features that the end user does not need while lacking features that are required.

In order to solve these issues, the author emphasizes that a better way in which to elicit requirements in agile development projects needs to be explored so as to develop cost-effective high-quality software that better meets users' expectations. The proposal presented consists in the introduction of collaborative Agile Usage-Centered Design sessions during the development process, during which developers, interaction designs, domain experts, business leaders, and the actual End-User participate in ``designing'' the requirements.

This study employs an experience report to provide evidence of the incorporation of Agile Usage-Centered Design into the day to day work of agile projects so as to deliver high quality software. It contributes with a detailed description of the experimental approach and offers a series of guidelines.

Based on the experience at Tomax Technologies, this approach appears to improve the stakeholders' understanding and ownership of the software. The paper also reports that priorities are easier to identify by looking into user roles and their focal task cases. In addition, Agile U-CD as an instance of Interaction Design is simple, teachable and repeatable. All these advantages make it possible to deliver high-quality solutions on time and better meet user expectations. As a counterpoint to these advantages it is necessary to consider that this approach requires constant collaboration, which can be exhausting.

The author comes to the conclusion that while agile development methods allow high quality software to be delivered sooner, adding interaction design concepts helps increase confidence and user empathy, which better leads to end-user satisfaction. Although the experience presented occurred within an extreme programming methodology, this approach also feeds other agile development methods well.

\subsubsection{Percival, Braz and Adilson}
\label{sec:PercivalBrazAdilson}

In [S6], the IBM Design Thinking method is presented. This method extends the original design thinking principles to agile software development by providing a new approach with which to capture requirements that includes end-user feedback during all the project development phases. The authors claim that traditional design thinking is insufficient to be integrated into an agile software development process. Software development demands a close relationship between designers and engineers, and traditional design thinking separates design from implementation.

Considering this, the presented study establishes two main objectives. Firstly, that of providing an easy description of the IBM Design Thinking method and its differences from traditional design thinking. Secondly, that of evaluating its benefits and limitations by means of a survey with the developers and designers of five real software development projects.

\begin{figure}[h]
	\includegraphics[width=\columnwidth]{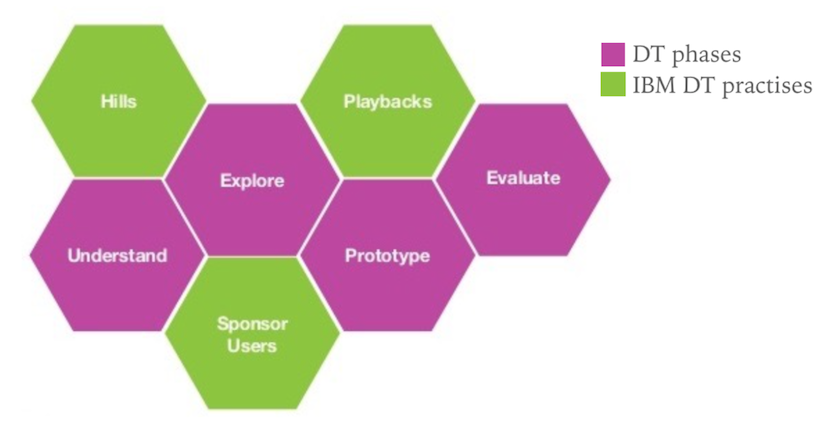}
	\caption{IBM DT compared to traditional DT [S6]}
	\label{fig:IBMDT}
\end{figure}

\begin{figure*}[h]
	\centering
	\includegraphics[width=0.8\textwidth]{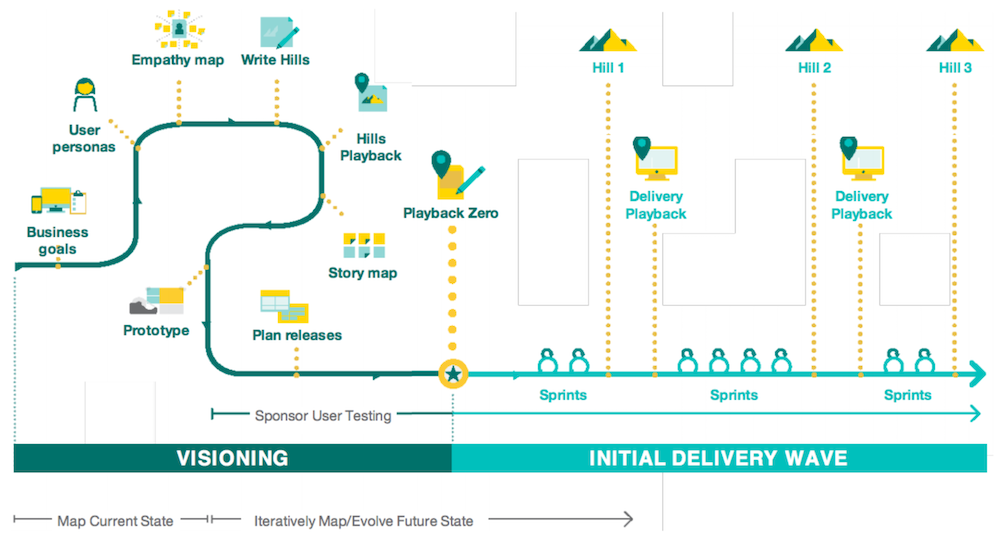}
	\caption{IBM DT Software Development Framework [S6]}
	\label{fig:IBMDT2}
\end{figure*}

This IBM Design Thinking Method introduces three new elements into the traditional design thinking process: hills, sponsor users and playbacks. Hills are a new way in which to express user's needs, sponsor users are real human beings who can share their experiences and points of view and playbacks are checkpoints when the project team and Sponsor Users meet in order to review the state of the project and plan the next steps. The comparison of this method with traditional design thinking is presented in Figure \ref{fig:IBMDT} and the complete extended method is shown in Figure \ref{fig:IBMDT2}.

The research methodology employed by this study is a survey that follows Barbara Kitchenham's guidelines for ``Personal Opinion Surveys'' \cite{kitchenham2008personal}. The results of the survey evidence that this method can help the development team to attain a better understanding of the problems that need to be solved and what the best solutions through which the user needs can be satisfied. This deep understanding becomes solutions that are valuable for the End-users. However, it is necessary to consider that, in order to apply this process, companies need to be willing to reorganize their teams, review their work model and functional roles and change their approach towards problem-solving.

The authors mention that this method could be used in incremental software development, such as cloud based software, and emphasize that further studies are needed to measure satisfaction levels accurately and to comprehend the limitations of the software framework developed.

\subsubsection{Lombriser, Lucassen and Brinkkemper}
\label{sec:LombriserLucassenBrinkkemper}

In [S7], the authors present the Gamified Requirements Engineering Model, which integrates gamification and engagement theories in the context of requirement engineering performance. The authors emphasize that despite the important role that requirements engineering plays in software development, insufficient stakeholder participation in requirement elicitation and poor collaboration and communication still persist as very common problems. This leads to low-quality and unsatisfactory requirements and consequently to the inability to deliver on time, within cost or expected scope.

The study aims, first, to improve the quality and increase creativity in requirements, and, second, to evaluate the effectiveness of gamification so as to improve stakeholder engagement and productivity in requirements engineering when online digital platforms are used. The GREM model consists in a relationship between three variables: gamification (defined as the application of game design elements in non-gaming contexts) \cite{deterding2011game}, engagement (defined as the emotional, cognitive and behavioral connection that exists, at any point in time and possibly over time, between a user and a resource) \cite{deci2010self,maslow1943theory} and performance.

\begin{figure}[h]
	\includegraphics[width=\columnwidth]{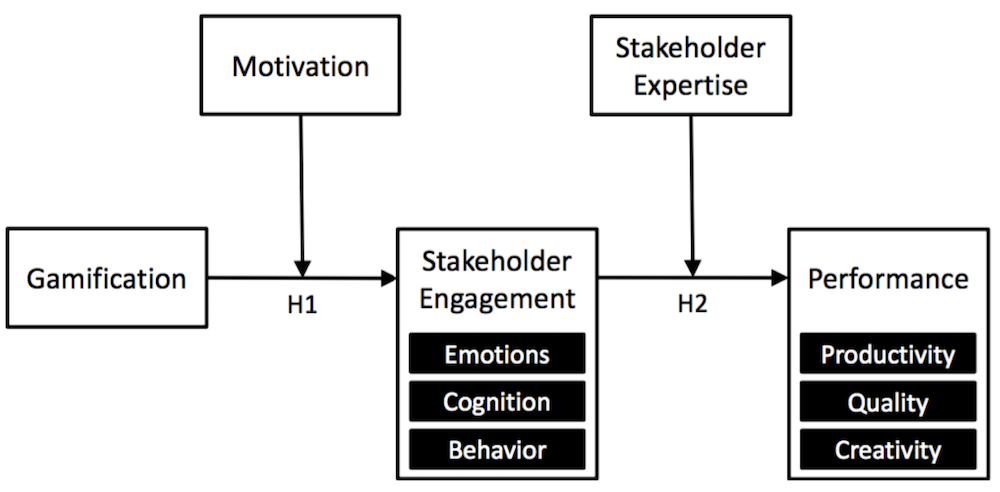}
	\caption{The Gamified Requirements Engineering Model [S7]}
	\label{fig:Gamified}
\end{figure}

Besides these three variables, two control variables are included in order to mitigate threats to internal validity: motivation and stakeholder expertise. Three sub-dimensions are defined for stakeholder engagement: emotions, cognition and behavior. Performance is sub-divided into productivity, quality and creativity, which are perceived as supportive concepts with which to measure the output in requirements elicitation. The relationships among all these concepts are shown in the conceptual model of Figure \ref{fig:Gamified}.

This work follows a controlled experiment research methodology. It documents an experiment that took place in a business environment (IT consultancy company in Munich, Germany) with a total of 12 stakeholders divided in two groups (control and experimental group) well equilibrated in terms of gender motivation and experience.

The results of the experiment evidenced this creativity improvement and performance increase in terms of the quantity and quality of the requirements. However, it was also observed that the success of gamification depends very much on the game elements and game mechanics. While rivalry elements seem to bring more creativity into requirements, they may not be suitable in phases during which cooperation and the exchange of ideas are needed. This leads the authors to a new hypothesis: ``While requirements elicitation is positively supported by competitive game elements, cooperative game elements are more suitable for requirements analysis, specification and validation''.

This leads to conclude that gamification can positively influence the elicitation process in agile requirements engineering. As future work, the authors emphasize that further experiments with different sample sizes and game mechanics are needed to generalize the results and the applicability of the GREM.

\subsubsection{Mahmud and Veneziano}
\label{sec:MahmudVeneziano}

In [S8], the authors present an approach with which can be used to elicit and represent requirements within the SCRUM model, based on mind-maps which establish the initial product backlog. Mind Mapping is a technique that encourages people to think of, organize and represent information within a radial hierarchy, by locating the most important concept at the center of a given diagram and relate it to other concepts situated farther away from the center of the diagram \cite{buzan2006mind}.

The motivation to pursue this study originated from the difficulties involved in combining traditional requirements elicitation with agile approaches, and from the problems derived from agile requirement elicitation, such as poor customer involvement or costly system changes owing to the anticipation of too many upfront requirements.

The authors propose to confront these issues by evaluating how the requirements elicitation process within agile software development projects, and in particular within the SCRUM model, may benefit from the use of mind maps to develop a suitable product backlog. This new technique could also reconcile traditional approaches in software development with agile methods.

The authors carry out an experiment to evaluate whether the quality of requirements represented as a backlog product in SCRUM could be affected by the adoption of mind-mapping techniques during their elicitation and analysis process. The results of the experiment show that the mind-mapping technique seems to support the analysis and capture of consistent and complete requirements very positively. However, it should be kept in mind that instructions and training on how to build mind-maps need to be provided to the stakeholders beforehand. Regarding the validity limits of the experiment, the results show that, by using a mind-mapping technique, the overall quality of the product backlog is significantly higher if the product owners have little experience and not worse if the product owners are more experienced. This leads the authors to conclude that mind-mapping and any derived technique could be used to set up the initial product backlog when developing by using agile methodologies like SCRUM. For future work, the authors suggest that more experiments and a larger amount of data are needed to obtain statistically significant results.

\subsubsection{Maiden}
\label{sec:Maiden}

In [S9], the author discusses the need to provide requirements analysts with a new creative and agile toolbox and the right instructions to use them. Although the focus is not on presenting a new creative technique or process for requirements engineering in agile development, these ideas are of great interest within the context of this systematic review. The author emphasizes that traditional elicitation tools, such as observations, interviews and questionnaires, have some weaknesses. These methods, which rely on communication, tend to generate a lot of information but still omit some information that the analyst needs because, for example, certain knowledge is taken for granted.

The article does not present empirical evidence but is based on the author's subjective experience and argumentative research. He encourages his readers to think about new and more effective ways in which to elicit, acquire, and discover customers' requirements. The article suggests that new creative techniques shared among stakeholders, analysts and developers may diminish the weakness of traditional elicitation tools.

The author does not specify any particular type of project or organization that could benefit from this new toolbox, but since he mentions requirements analysts, it can be deduced that this new toolbox could be used on a wide variety of software projects.

The author concludes that analysts and stakeholders should look to agile development techniques and user-centered design for techniques such as the analysis of Web analytics, wireframing, and user stories and exchange their experiences and techniques with each other, not via paper, but using social media.

\subsubsection{Wanderley, Silveire and Araujo}
\label{sec:WanderleySilveireAraujo}

The proposal presented in [S10] aims to give support to the modeling of requirements by means of a creative and agile technique based on the automatic transformation of mind maps into feature models based on model-driven engineering (MDE). The authors emphasize that the formalization of requirements through modeling tools requires a technical knowledge that not all domain experts already have. Furthermore, many system requirements are not well understood, which increases the difficulty involved in producing conceptual models efficiently. On top of that, communication problems between domain experts and software engineers are very common.

\begin{figure}[h]
	\includegraphics[width=\columnwidth]{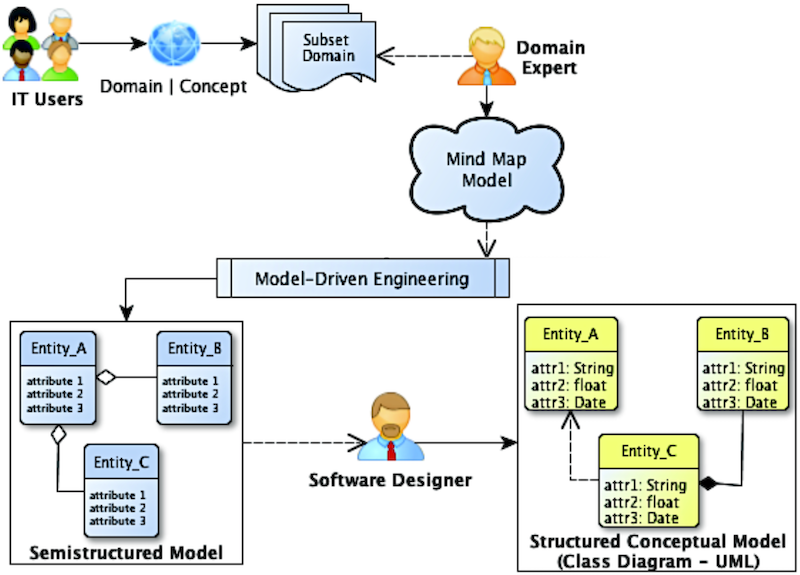}
	\caption{Solution proposed by [S12]}
	\label{fig:Solution}
\end{figure}

\begin{figure}[h]
	\includegraphics[width=\columnwidth]{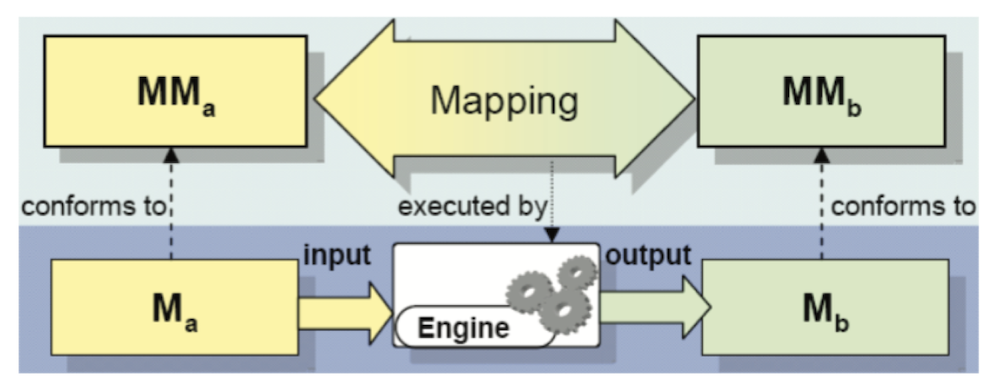}
	\caption{Diagram of transformation among Models [S10]}
	\label{fig:DiagramTrans}
\end{figure}

\begin{figure}[h]
	\includegraphics[width=\columnwidth]{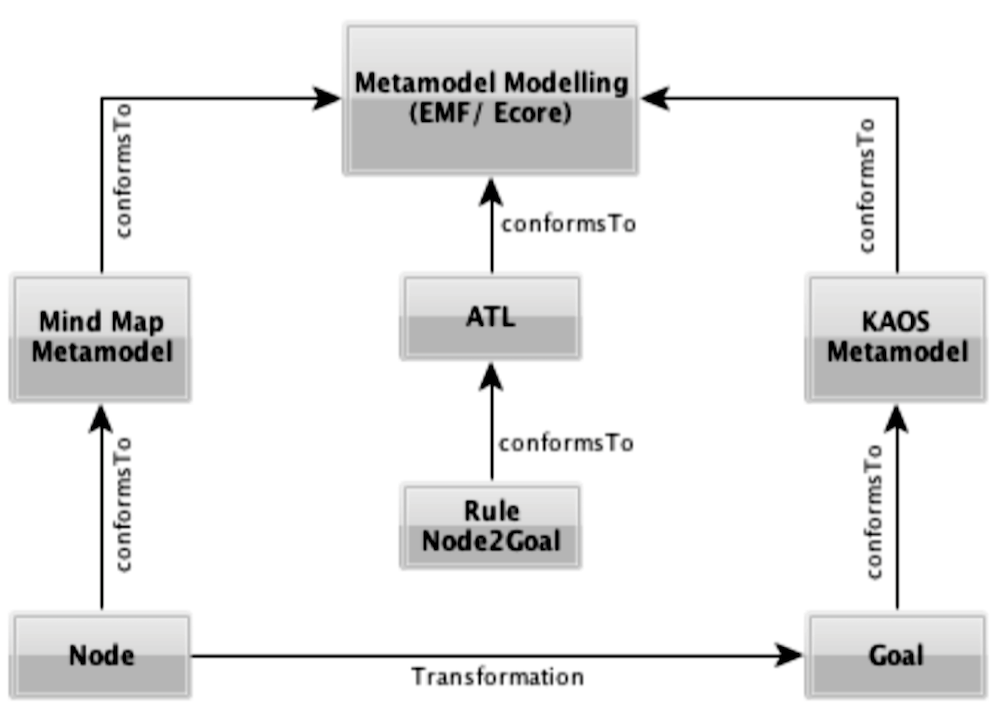}
	\caption{Transformation process among models [S11]}
	\label{fig:TransModels}
\end{figure}

All these reasons motivate the authors to present an approach that could facilitate and improve the domain analysis and modeling process. The main contributions of this work are the definition of a modeling environment that supports the creation of feature models by domain experts, regardless of their knowledge of requirements engineering. An overview of the proposed solution can be seen in Figures \ref{fig:Solution}, \ref{fig:DiagramTrans} and \ref{fig:TransModels}. In this study, the authors present an example of the approach with the tool TaRGet, whose objective is to automatically generate a suite of tests based on the specifications of cases written in natural language, but they do not provide any empirical evidence.

The same authors continue this line of investigation in [S11], in which they present a study that aims to evaluate the use of creative requirements models together with goal-oriented models in a model-driven context. The approach establishes the mapping between the main concepts of the KAOS framework (which stands for ``Keep All Objects Satisfied'') and mind map models.

In a subsequent study [S12], the authors provide empirical evidence of the use of the initial model [S10] with two controlled experiments that involve senior, middle and junior software designers from industry and academia. The objective of these experiments is to evaluate the effort needed to build conceptual models expressed as UML class diagrams with the support of mind maps and to evaluate qualitatively the effectiveness of a mind map as regards understanding the domain analyzed.

The results of the experiments show that this approach simplifies the modeling process in an agile and creative manner by using mind maps, and facilitates the negotiation process related to the variability analysis of a domain. In other words, less effort is required to produce conceptual models through the use of mind maps than without them. However, in terms of productivity, the difference between the time required when using mind maps and when not using them is not meaningful.

The study also proves that it is possible to transform mind maps into feature models, without the need to refine the final model. This makes it possible to deduce that mind maps, with the extensions defined in their metamodels, can be used as an agile tool with which to model features. As future work, the authors propose to extend and formalize the transformation rules with a specific ATL transformation language.

\subsubsection{Bones, Harrison and Liu}
\label{sec:BonesHarrisonLiu}

In [S13], [S14] and [S15] the authors describe and investigate the use of a creative technique that can be used together with agile programming for requirements engineering. The proposal is based on disciplined goal-responsibility modeling but introduces a non-formal and creative method with which to produce goal responsibility models under agile constraints: time, incompleteness and catching up after an initial creative burst.

Goal responsibility models represent the stakeholders' intentions for a system-to-be that will operate in an expected environment, in fulfillment of a contract. These models, when created in the very early phases of the project by combining all objectives and behaviors, are able to create realistic expectations in the minds of the stakeholders and are helpful when appraising the model for its feasibility, adequacy and testability. However, the creation of appropriate goal models can be complex when combined with agile constraints. This is particularly the case when stakeholders express their requirements as partial, hypothetical or functional designs, and in backlog driven projects which, after a few sprints, reach the point at which there are inadequate specifications for regression and acceptance testing.

These difficulties motivate the authors to investigate new creative techniques that can be used to produce goal responsibility models. In [S13], they describe goal sketching, a technique based on natural language that consists in creating a goal graph which expresses the high-level motivations behind the intention to develop the software. This initial graph is refined throughout a number of stages in a manner similar to the use of Scrum sprints \cite{rising2000scrum}. During the refinement process, the goals are refined only as necessary for the current stage, which is called lazy refinement. The graph that is present during the process is called a ``stage graph''.

In a subsequent study [S14], the authors explore the use of UML diagrams such as activity diagrams (which are good at representing functionality and processes) to accelerate the speed of sketch drafting and refactoring. They validate the proposal by employing an action research methodology and testing the technique in both small and large development projects, such as the development of a product supported by venture capital, the specification, procurement and acceptance of a management information system for a food processing company and a tool to support professional services in healthcare.

Although the data obtained from the presented cases is not very significant, the results show that this technique can adapt to changing requirements and cope with unresolved requirements. This leads the authors to conclude that this technique enables a successful development of evolvable systems and is in compliant with the real world needs of industrial software development.

Future work includes applying the method to more industrial projects, developing tools to accelerate sketching and refactoring, and exporting goal sketches into KAOS, or generalizing the method to enable the incorporation of other structured and UML modeling methods.

\subsubsection{Hastreiter, Krause, Schneidermeier and Wolff}
\label{sec:HastreiterKrauseSchneidermeierWolff}

In [S16], the authors propose a collaborative tool for mobile prototyping that enables the user to sketch mobile applications with a set of basic, customizable objects and to share their prototypes with the rest of the team. After performing an evaluation of existing sketching and prototyping tools for mobile application platforms by means of a competitive analysis, the authors highlight that, despite the importance of usability in aspects such as the efficiency of the system, effectiveness in the work process or the user's satisfaction, current software is not sufficiently attractive for a broader audience. Furthermore, existing mobile solutions do not support collaboration, which has been emphasized as an essential part of the development process. The analysis also states that the creation of complex interactive high-fidelity prototypes is not practical in a mobile work setting.

These problems motivate the authors to develop a requirements elicitation approach focused on UX methods to support action mode usage in order to leverage creative potentials.

\begin{figure}[h]
	\includegraphics[width=\columnwidth]{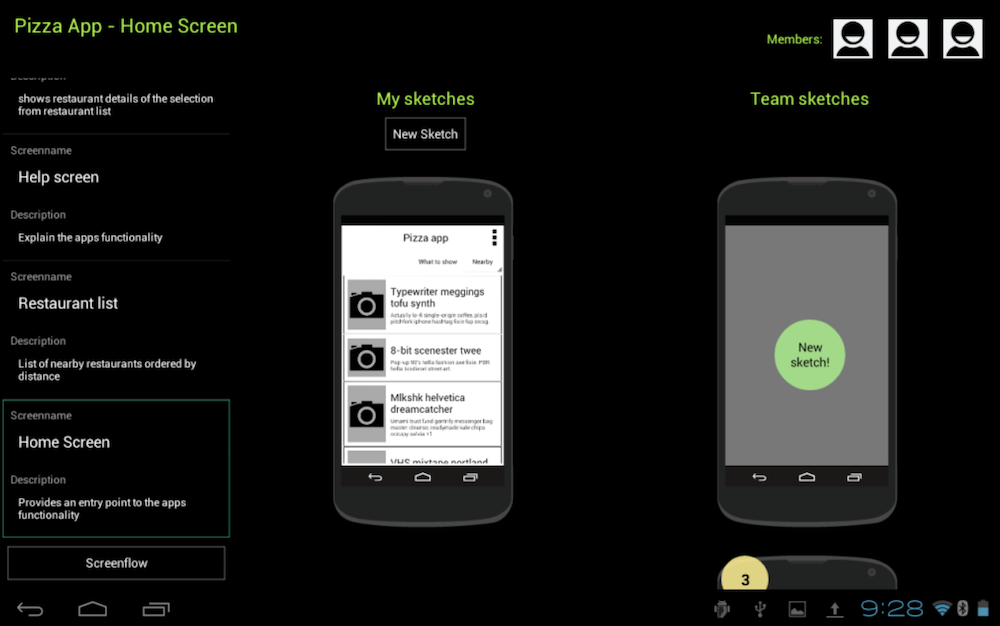}
	\caption{Overview of the Prime platform [S17]}
	\label{fig:Platform}
\end{figure}

The ``Prime'' proposal consists of a prototyping tool for android $10\,''$ tablets that offers a low fidelity prototyping functionality (see Figure \ref{fig:Platform}). The interface is composed of a drawing area into which objects can be dragged and dropped from a palette and be directly manipulated to fit the users' needs. Sketches and drawings can be incorporated into the prototype by means of the hardware camera. In order to enable collaboration between users, the tool includes an online backend based on parse.com (Parse Platform). This cloud platform provides easy to set up user accounts with different privileges and a database in which to store arbitrary objects.

During the development process, the authors apply different UX techniques among small and multidisciplinary development teams practicing agile methods. One of them is SHIRA, ``Structured Hierarchical Interview for Requirement Analysis'', an interviewing technique that seeks to explore the meaning of abstract product qualities, such as ``controllable'', ``simple'', ``impressive'' or ``innovative'', for a specific software product in a specific context of use.

The results of the study show that this tool can speed up the development process and provide an adequate solution for the prototyping of mobile applications. It also improves user collaboration by allowing the users to share their prototypes with the rest of the team and maintains all team members updated. As future work, the authors suggest the further optimization of the tool and the design of a method set with which to emphasize requirements and to evaluate ideas during the engineering process.

\subsubsection{Sulmon, Derboven, Montero and Zaman}
\label{sec:SulmonDerbovenMonteroZaman}

In [S17], the authors propose the User-Driven Creativity Framework, a participatory design methodology based on creativity theories that aims to stimulate end user creativity in the early design process of innovative applications with which to elicit user requirements. The authors highlight that agile methods tend to assume that users can easily say what they want if asked and come up with innovative ideas. However, the reality is that gathering requirements takes lots of time and effort. Although it is claimed that agile methodologies promote collaboration between the customer and the development team, very few define how the people on the team that play the role of customer can learn what the real end user needs and how they can accurately represent those needs. This is the case of, for example, Scrum projects, in which the product owner is the person who represents the end user.

\begin{figure}[h]
	\includegraphics[width=\columnwidth]{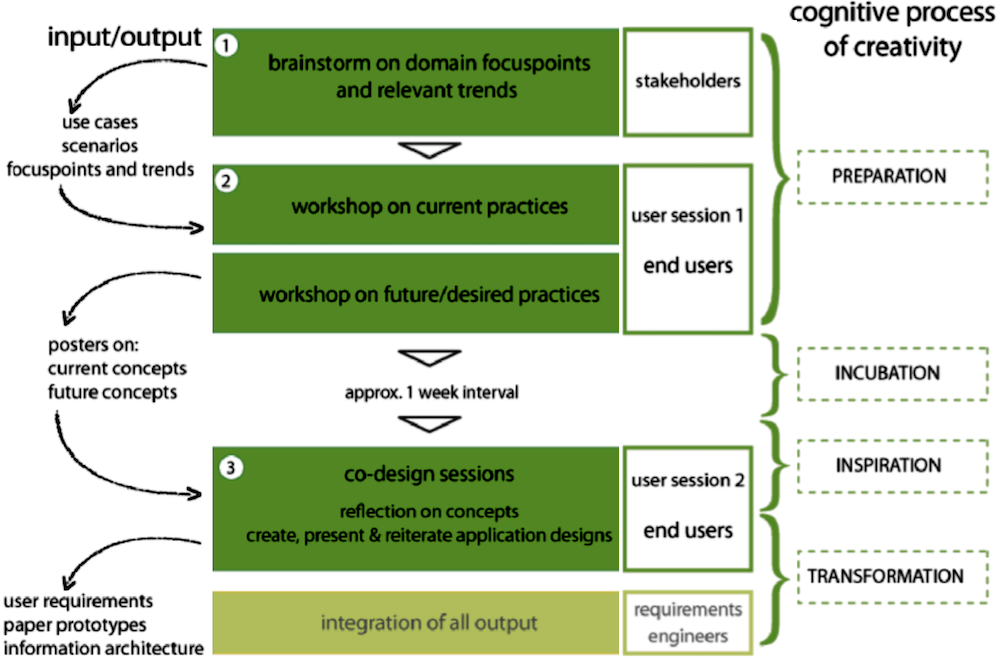}
	\caption{The User Driven Creativity Framework [S17]}
	\label{fig:UserCreatvity}
\end{figure}

These problems motivate the authors to develop a framework as a viable alternative to enable requirements engineers or customer representatives to efficiently attain active stakeholder participation for initial requirements gathering within a limited amount of time. It consists of a 4-creative stage process, represented in Figure \ref{fig:UserCreatvity}.

\begin{enumerate}
	\item \textbf{Preparation Stage}: this consists of two activities. Firstly, an initial brainstorming activity with subject matter experts and other stakeholders serves to gather as much background information as possible. A session with actual end users then takes place. During this session, users are encouraged to actively engage with each other in small group discussions and analyze the relevant information. The session is divided into two parts, the first of which focuses on current practices in the domain, and the second of which focuses on the desired future practices. 
	\item \textbf{Incubation Stage}: this consists of leaving a break of at least three days between the first and second user sessions. This relaxation time offers participants ``room for thought'' and time to process the outcome of the first session.
	\item \textbf{Inspiration Stage}: this is a co-design session, in which participants collaborate to create paper prototypes of the ideas gathered during the first user session. The results from this session can be used to understand users' needs and values and help designers and professional to finalize the design process.
	\item \textbf{Transformation Stage}: this is the phase in which all the prototypes and information gathered are formally integrated by requirements engineers. 
\end{enumerate}

The results of the case study were rich and elaborate, and obtained in a limited amount of time. This proves that this framework can be successfully used to unveil high-level requirements and effectively yield the in-depth user knowledge and involvement required to establish a strong foundation for further agile development activities. For future work, the authors underline the need for a refinement and an in-depth evaluation of the method. They also suggest that this framework could be applied in the design processes of other new application domains, in which a creativity-focused, user-oriented approach is appropriate.

\begin{table*}[htb]
	\centering
	\caption{Synthesis of RQ1}
	\label{tab:SyntRQ1}
	\resizebox{\textwidth}{!}{%
		\begin{tabular}{lcc|ccccccccccccccccc}
			\toprule
			\multicolumn{20}{l}{\textit{RQ1: What methodologies or techniques that involve creative approaches in requirements elicitation (RE) in agile software development (ASD) exist?}} \\ \midrule \addlinespace
			& \multicolumn{2}{c}{\textbf{VALUES}} & \multicolumn{17}{c}{\textbf{SYNTHESIS}} \\ \cline{2-20} \addlinespace
			& \textbf{1} & \textbf{0 }& \textbf{S1} & \textbf{S2} & \textbf{S3} & \textbf{S4} & \textbf{S5} & \textbf{S6} & \textbf{S7} & \textbf{S8} & \textbf{S9} & \textbf{S10} & \textbf{S11} & \textbf{S12} & \textbf{S13} & \textbf{S14} & \textbf{S15} & \textbf{S16} & \textbf{S17} \\ \addlinespace
			\textbf{Process} & YES & NO & 1 & 1 & 0 & 1 & 0 & 0 & 0 & 0 & 0 & 0 & 0 & 0 & 0 & 0 & 0 & 0 & 0 \\
			\textbf{Framework} & YES & NO & 0 & 0 & 0 & 0 & 0 & 1 & 0 & 0 & 0 & 0 & 0 & 0 & 0 & 0 & 0 & 0 & 1 \\
			\textbf{Tools} & YES & NO & 0 & 0 & 0 & 0 & 0 & 0 & 0 & 0 & 1 & 0 & 0 & 0 & 0 & 0 & 0 & 1 & 0 \\
			\textbf{Technique} & YES & NO & 0 & 0 & 1 & 0 & 1 & 0 & 0 & 1 & 0 & 1 & 1 & 1 & 1 & 1 & 1 & 0 & 0 \\
			\textbf{Model} & YES & NO & 0 & 0 & 0 & 0 & 0 & 0 & 1 & 0 & 0 & 0 & 0 & 0 & 0 & 0 & 0 & 0 & 0 \\
			\bottomrule
		\end{tabular}%
	}
\end{table*}

\begin{table*}[htb]
	\centering
	\caption{Synthesis of RQ2}
	\label{tab:SyntRQ2}
	\resizebox{\textwidth}{!}{%
		\begin{tabular}{lcc|ccccccccccccccccc}
			\toprule
			\multicolumn{20}{l}{\textit{RQ2: What empirical evidence is there of the application of creative techniques for RE in ASD?}} \\ \midrule \addlinespace
			& \multicolumn{2}{c}{\textbf{VALUES}} & \multicolumn{17}{c}{\textbf{SYNTHESIS}} \\ \cline{2-20} \addlinespace
			& \textbf{1} & \textbf{0 }& \textbf{S1} & \textbf{S2} & \textbf{S3} & \textbf{S4} & \textbf{S5} & \textbf{S6} & \textbf{S7} & \textbf{S8} & \textbf{S9} & \textbf{S10} & \textbf{S11} & \textbf{S12} & \textbf{S13} & \textbf{S14} & \textbf{S15} & \textbf{S16} & \textbf{S17} \\ \addlinespace
			
			\textbf{Experiment} & YES & NO & 1 & 0 & 0 & 0 & 0 & 0 & 1 & 1 & 0 & 0 & 0 & 0 & 0 & 0 & 0 & 0 & 0 \\
			\textbf{Action Research} & YES & NO & 0 & 1 & 0 & 0 & 0 & 0 & 0 & 0 & 0 & 0 & 0 & 0 & 1 & 1 & 1 & 1 & 0 \\
			\textbf{Case of study} & YES & NO & 0 & 0 & 0 & 1 & 0 & 0 & 0 & 0 & 0 & 1 & 1 & 0 & 0 & 0 & 0 & 0 & 1 \\
			\textbf{Experience Report} & YES & NO & 0 & 0 & 0 & 0 & 1 & 0 & 0 & 0 & 1 & 0 & 0 & 0 & 0 & 0 & 0 & 0 & 0 \\
			\textbf{Survey} & YES & NO & 0 & 0 & 0 & 0 & 0 & 1 & 0 & 0 & 0 & 0 & 0 & 0 & 0 & 0 & 0 & 0 & 0 \\
			
			\bottomrule
		\end{tabular}%
	}
\end{table*}

\begin{table*}[htb]
	\centering
	\caption{Synthesis of RQ3}
	\label{tab:SyntRQ3}
	\resizebox{\textwidth}{!}{%
		\begin{tabular}{lcc|ccccccccccccccccc}
			\toprule
			\multicolumn{20}{l}{\textit{RQ3: What is known about the benefits and limitations of using creative approaches for RE in ASD?}} \\ \midrule \addlinespace
			& \multicolumn{2}{c}{\textbf{VALUES}} & \multicolumn{17}{c}{\textbf{SYNTHESIS}} \\ \cline{2-20} \addlinespace
			
			\textbf{BENEFITS} & \textbf{1} & \textbf{0 }& \textbf{S1} & \textbf{S2} & \textbf{S3} & \textbf{S4} & \textbf{S5} & \textbf{S6} & \textbf{S7} & \textbf{S8} & \textbf{S9} & \textbf{S10} & \textbf{S11} & \textbf{S12} & \textbf{S13} & \textbf{S14} & \textbf{S15} & \textbf{S16} & \textbf{S17} \\ \addlinespace

			\textbf{Improvement of stakeholder engagement} & YES & NO & 0 & 0 & 0 & 0 & 1 & 0 & 0 & 0 & 0 & 1 & 1 & 1 & 0 & 0 & 0 & 1 & 1 \\
			\textbf{Improvements in communication and ideas interchange} & YES & NO & 0 & 1 & 0 & 1 & 0 & 0 & 0 & 0 & 0 & 0 & 0 & 0 & 0 & 0 & 0 & 1 & 1 \\
			\textbf{Improvement in novelty, quality and usefulness of rqmts} & YES & NO & 1 & 0 & 0 & 0 & 0 & 0 & 1 & 0 & 0 & 0 & 0 & 0 & 0 & 0 & 0 & 0 & 1 \\
			\textbf{Better meet business needs}  & YES & NO & 0 & 1 & 0 & 0 & 1 & 1 & 0 & 0 & 1 & 1 & 1 & 1 & 1 & 1 & 1 & 1 & 0 \\
			\textbf{Ease problem understanding and rqmts elicitation} & YES & NO & 0 & 0 & 1 & 1 & 1 & 1 & 0 & 0 & 1 & 1 & 1 & 1 & 1 & 1 & 1 & 1 & 1 \\ \addlinespace \addlinespace

			\textbf{LIMITATIONS} & \textbf{1} & \textbf{0 }& \textbf{S1} & \textbf{S2} & \textbf{S3} & \textbf{S4} & \textbf{S5} & \textbf{S6} & \textbf{S7} & \textbf{S8} & \textbf{S9} & \textbf{S10} & \textbf{S11} & \textbf{S12} & \textbf{S13} & \textbf{S14} & \textbf{S15} & \textbf{S16} & \textbf{S17} \\ \addlinespace

			\textbf{Requirements are seen as less useful} & YES & NO & 1 & 0 & 0 & 0 & 0 & 0 & 0 & 0 & 0 & 0 & 0 & 0 & 0 & 0 & 0 & 0 & 0 \\
			\textbf{Users must be open-minded and willing to collaborate} & YES & NO & 0 & 0 & 0 & 1 & 1 & 0 & 0 & 0 & 0 & 0 & 0 & 0 & 0 & 0 & 0 & 1 & 1 \\
			\textbf{Reorganization of teams might be needed} & YES & NO & 0 & 0 & 0 & 0 & 0 & 1 & 0 & 0 & 0 & 0 & 0 & 0 & 0 & 0 & 0 & 0 & 0 \\
			\textbf{Training in novel techniques or process is needed} & YES & NO & 1 & 1 & 1 & 1 & 1 & 1 & 0 & 1 & 0 & 1 & 1 & 1 & 1 & 1 & 1 & 0 & 0 \\
			
			\bottomrule
		\end{tabular}%
	}
\end{table*}

\begin{table*}[htb]
	\centering
	\caption{Synthesis of RQ4}
	\label{tab:SyntRQ4}
	\resizebox{\textwidth}{!}{%
		\begin{tabular}{llccc|ccccccccccccccccc}
			\toprule
			\multicolumn{22}{l}{\textit{RQ4. In what sort of organizations or projects using ASD could the use of these creative approaches be most suitable?}} \\ \midrule \addlinespace
			
			& & \multicolumn{3}{c}{\textbf{VALUES}} & \multicolumn{17}{c}{\textbf{SYNTHESIS}} \\ \cline{3-22} \addlinespace
			
			& & \textbf{1} & \textbf{0}& \textbf{-1} & \textbf{S1} & \textbf{S2} & \textbf{S3} & \textbf{S4} & \textbf{S5} & \textbf{S6} & \textbf{S7} & \textbf{S8} & \textbf{S9} & \textbf{S10} & \textbf{S11} & \textbf{S12} & \textbf{S13} & \textbf{S14} & \textbf{S15} & \textbf{S16} & \textbf{S17} \\ \addlinespace
			
			\multirow{2}{*}{\textbf{Organization size}} & \textbf{Large Organization} & YES & NS & NO & 1 & 1 & 0 & 1 & 1 & 1 & 0 & 0 & 0 & 1 & 1 & 1 & 1 & 1 & 1 & 0 & 0 \\
			& \textbf{Startup or Small organization} & YES & NS & NO & 0 & 0 & 0 & 0 & 0 & 0 & 0 & 0 & 0 & 1 & 1 & 1 & 1 & 1 & 1 & 1 & 0 \\ \addlinespace
			\multirow{2}{*}{\textbf{Duration}} & \textbf{Short development projects} & YES & NS & NO & 0 & 0 & 0 & 0 & 0 & -1 & 0 & 0 & 0 & 0 & 0 & 0 & 1 & 1 & 1 & 1 & 0 \\
			& \textbf{Long-term development projects} & YES & NS & NO & 0 & 1 & 0 & 1 & 1 & 1 & 0 & 0 & 0 & 0 & 0 & 0 & 0 & 0 & 0 & 0 & 0 \\ \addlinespace
			\multirow{4}{*}{\textbf{Type of development}} &	\textbf{Mobile Applications} & YES & NS & NO & 0 & 0 & 0 & 0 & 0 & 1 & 0 & 0 & 0 & 1 & 1 & 1 & 0 & 0 & 0 & 1 & 1 \\
			& \textbf{Website development} & YES & NS & NO & 1 & 0 & 0 & 0 & 0 & 0 & 0 & 0 & 0 & 0 & 0 & 0 & 0 & 0 & 0 & 0 & 0 \\
			& \textbf{Enterprise architecture} & YES & NS & NO & 0 & 0 & 1 & 0 & 0 & 1 & 0 & 0 & 0 & 1 & 0 & 0 & 1 & 1 & 1 & 0 & 0 \\
			& \textbf{Web application} & YES & NS & NO & 0 & 0 & 0 & 1 & 0 & 1 & 0 & 0 & 0 & 0 & 1 & 1 & 1 & 1 & 1 & 0 & 0 \\
			
			\bottomrule
		\end{tabular}%
	}
\end{table*}

\subsection{Classification of studies}
\label{sec:ClassificationStudies}

During the review and extraction stage, we defined a table to synthesize the data gathered to answer each research question. The aim was to help to quantify the qualitative information extracted from each study. To that end, each study is assigned a value for each research question.

Table \ref{tab:SyntRQ1} summarizes the results obtained for RQ1. For instance, 52\% of studies reviewed propose some technique to leverage creativity in requirements elicitation.

Table \ref{tab:SyntRQ2} shows the data gathered to answer RQ2. It is worth noting that there is not preferred method since the most used is action research, which is adopted only by 29\% of the selected studies.

Table \ref{tab:SyntRQ3} quantitatively synthesizes the data gathered with regard to strengths and limitations of creative approaches for Requirements Engineering in ASD. The most outstanding finding is that these proposals ease problem-understanding according to 76\% of the studies selected.

Table \ref{tab:SyntRQ4} focuses on the data gathered regarding the most suitable scenarios to apply these proposals. The fact that most of the reviewed studies deployed their proposals in large organizations clearly stands out (64\%).

Finally, Table \ref{tab:SyntRQ5} synthesizes the data related with RQ5, showing that there is no interest in a particular agile methodology or method. The most popular one being agile model-driven development, which is only adopted by 23\% of the works selected.

\begin{table*}[htb]
	\centering
	\caption{Synthesis of RQ5}
	\label{tab:SyntRQ5}
	\resizebox{\textwidth}{!}{%
		\begin{tabular}{lcc|ccccccccccccccccc}
			\toprule
			\multicolumn{20}{l}{\textit{RQ5. Which agile methodologies are being used to integrate creative techniques into agile software development RE processes?}} \\ \midrule \addlinespace
			& \multicolumn{2}{c}{\textbf{VALUES}} & \multicolumn{17}{c}{\textbf{SYNTHESIS}} \\ \cline{2-20} \addlinespace
			& \textbf{1} & \textbf{0 }& \textbf{S1} & \textbf{S2} & \textbf{S3} & \textbf{S4} & \textbf{S5} & \textbf{S6} & \textbf{S7} & \textbf{S8} & \textbf{S9} & \textbf{S10} & \textbf{S11} & \textbf{S12} & \textbf{S13} & \textbf{S14} & \textbf{S15} & \textbf{S16} & \textbf{S17} \\ \addlinespace
			\textbf{SCRUM} & YES & NO & 1 & 0 & 0 & 0 & 0 & 0 & 0 & 1 & 0 & 0 & 0 & 0 & 0 & 0 & 0 & 0 & 0 \\
			\textbf{Extreme Programming} & YES & NO & 0 & 0 & 0 & 0 & 1 & 0 & 0 & 0 & 0 & 0 & 0 & 0 & 0 & 0 & 0 & 0 & 0 \\
			\textbf{Agile Model Driven Development*} & YES & NO & 0 & 0 & 0 & 0 & 0 & 0 & 0 & 0 & 0 & 1 & 1 & 1 & 0 & 0 & 0 & 0 & 1 \\
			\bottomrule
			\multicolumn{20}{r}{\textit{{\tiny *MDE itself is not an agile methodology, here is being considered as an application to agile for rapid conversion of the mind maps to formal models}}}
		\end{tabular}%
	}
\end{table*}

\subsection{Quality assessment}
\label{sec:QualityAssessment}

The results obtained after applying the quality criteria to each of the selected studies are shown in Table \ref{tab:OverviewQA} and Table \ref{tab:OverviewSQA}.

In terms of rigor, we found that 15 out of 17 (88\%) studies obtained the maximum grade. Since most of the studies present very new approaches, it is in their self-interest to provide a clear description of the aims of the study (QA1), their approach (QA2) and follow a systematic methodology (QA3), thus enabling other researchers or practitioners to test their proposals.

In terms of credibility, 8 studies (47\%) qualified with the maximum grade. This represents less than half of the selected studies. We found that when documenting the findings (QA4) and the research process (QA5), some of the works did not provide any type of graphic representation that could facilitate an understanding of the results, or that some parts of the research process were omitted. With reference to validation (QA6), owing to the innovative nature of the approaches and the field of study, it was to be expected that not many studies would obtain the highest grade. A high percentage of the studies were either still testing their approaches or the data provided was not significant.

To conclude our quality assessment, we assessed the studies in terms of relevance. In this regard, 9 studies (53\%) obtained the highest grade. As for the extension of knowledge (QA8), all the studies obtained the maximum grade, since they proposed a new approach or extended understanding. With reference to the number of citations (QA9) it is not surprising that, owing to the innovative nature of the proposals they were not, on average, cited by many authors. In order to evaluate QA9, we carried out a Google Scholar search and counted the number of citations on the 20\textsuperscript{th} of June of 2017. Finally, we evaluated whether the proposals could be replicated in other organization or settings, and most of the studies obtained a high grade in this respect.

\begin{table*}[htb]
	\centering
	\caption{Overview of quality assessment}
	\label{tab:OverviewQA}
	\resizebox{\textwidth}{!}{%
		\begin{tabular}{llccccccccccccccccc}
			\toprule
			\textbf{Category} & \textbf{Item}& \textbf{S1} & \textbf{S2} & \textbf{S3} & \textbf{S4} & \textbf{S5} & \textbf{S6} & \textbf{S7} & \textbf{S8} & \textbf{S9} & \textbf{S10} & \textbf{S11} & \textbf{S12} & \textbf{S13} & \textbf{S14} & \textbf{S15} & \textbf{S16} & \textbf{S17} \\ \addlinespace

			\multirow{3}{*}{\textbf{Rigour}} & QA1 & 2 & 2 & 2 & 2 & 2 & 2 & 2 & 2 & 0 & 2 & 2 & 2 & 2 & 2 & 2 & 2 & 2 \\
			
			& QA2 & 2 & 2 & 2 & 2 & 2 & 2 & 2 & 2 & 0 & 2 & 2 & 2 & 2 & 2 & 2 & 2 & 2 \\
			
			& QA3 & 2 & 1 & 2 & 2 & 2 & 2 & 2 & 2 & 0 & 2 & 2 & 2 & 2 & 2 & 2 & 2 & 2 \\ \cline{3-19} \addlinespace

			& & 6 & 5 & 6 & 6 & 6 & 6 & 6 & 6 & 0 & 6 & 6 & 6 & 6 & 6 & 6 & 6 & 6 \\ \addlinespace \addlinespace
			
			\multirow{4}{*}{\textbf{Credibility}}  & QA4 & 1 & 1 & 2 & 2 & 2 & 2 & 2 & 2 & 2 & 2 & 2 & 2 & 2 & 2 & 2 & 2 & 2 \\
			
			& QA5 & 2 & 1 & 2 & 2 & 2 & 2 & 2 & 2 & 0 & 2 & 2 & 2 & 2 & 2 & 2 & 2 & 2 \\
			
			& QA6 & 2 & 2 & 0 & 2 & 1 & 2 & 2 & 1 & 0 & 1 & 1 & 2 & 1 & 2 & 2 & 2 & 2 \\
			
			& QA7 & 2 & 1 & 2 & 2 & 2 & 2 & 2 & 2 & 2 & 2 & 2 & 2 & 2 & 2 & 2 & 2 & 2 \\ \cline{3-19} \addlinespace

			& & 7 & 5 & 6 & 8 & 7 & 8 & 8 & 7 & 4 & 7 & 7 & 8 & 7 & 8 & 8 & 8 & 8 \\ \addlinespace \addlinespace

			\multirow{3}{*}{\textbf{Relevance}} & QA8 & 2 & 2 & 2 & 2 & 2 & 2 & 2 & 2 & 2 & 2 & 2 & 2 & 2 & 2 & 2 & 2 & 2 \\
			
			& QA9 & 2 & 0 & 0 & 2 & 2 & 0 & 2 & 2 & 0 & 2 & 2 & 1 & 2 & 2 & 1 & 1 & 1 \\
			
			& QA10 & 2 & 1 & 2 & 2 & 2 & 2 & 2 & 2 & 1 & 2 & 2 & 2 & 2 & 2 & 2 & 2 & 2 \\ \cline{3-19} \addlinespace

			& & 6 & 3 & 4 & 6 & 6 & 4 & 6 & 6 & 3 & 6 & 6 & 5 & 6 & 6 & 5 & 5 & 5 \\ \addlinespace \addlinespace \addlinespace

			& \textbf{Total} & \textbf{19} & \textbf{13} & \textbf{16} & \textbf{20} & \textbf{19} & \textbf{18} & \textbf{20} & \textbf{19} & \textbf{7} & \textbf{19} & \textbf{19} & \textbf{19} & \textbf{19} & \textbf{20} & \textbf{19} & \textbf{19} & \textbf{19} \\
			\bottomrule
		\end{tabular}%
	}
\end{table*}

\begin{table*}[htb]
	\centering
	\caption{Overview of studies per quality assessment}
	\label{tab:OverviewSQA}
	\resizebox{\textwidth}{!}{%
		\begin{tabular}{lccccccccccccccccc}
			\toprule
			\textbf{Study} & S4 & S7 & S14 & S1 & S5 & S8 & S10 & S11 & S12 & S13 & S15 & S16 & S17 & S6 & S3 & S2 & S9 \\ \addlinespace
			\textbf{Qualification} & 20 & 20 & 20 & 19 & 19 & 19 & 19 & 19 & 19 & 19 & 19 & 19 & 19 & 18 & 16 & 13 & 7 \\
			\bottomrule
		\end{tabular}%
	}
\end{table*}

\subsection{Publications}
\label{sec:Publications}

This section provides a brief analysis of the studies in terms of year, country and type of publication.

\subsubsection{Publication distribution per year}
\label{sec:PublicationYear}

\begin{figure}[h]
	\includegraphics[width=\columnwidth]{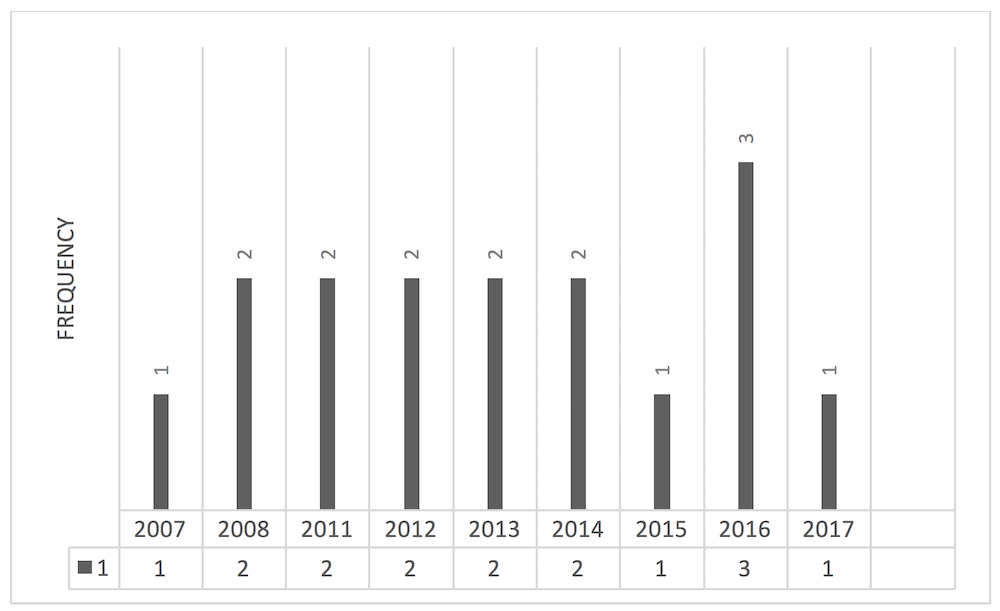}
	\caption{Distribution per year of primary studies}
	\label{fig:DistributionYear}
\end{figure}

Figure \ref{fig:DistributionYear} shows the distribution of the selected studies per year. Since the number of selected primary studies is not large, it is difficult to identify trends. We can, however, observe that all the selected studies can be found within the period of the last 15 years, which gives evidence of the innovative nature of the research field.

\subsubsection{Publication distribution per country and continent}
\label{sec:PublicationCountry}

\begin{figure}[h]
	\includegraphics[width=\columnwidth]{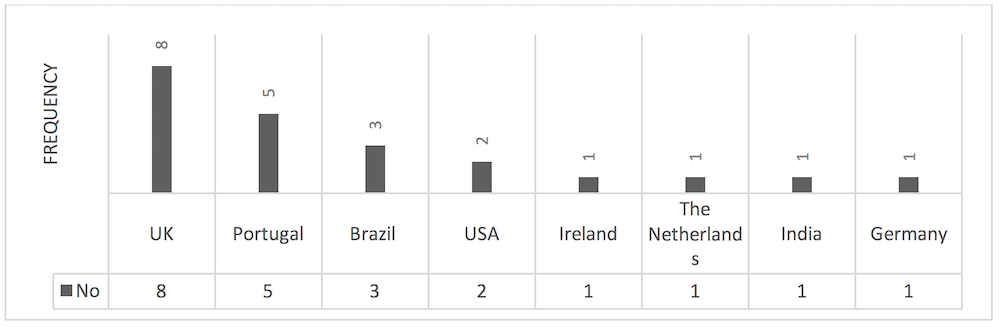}
	\caption{Publication distribution per country of primary studies}
	\label{fig:DistributionCountry}
\end{figure}

\begin{figure}[h]
	\includegraphics[width=\columnwidth]{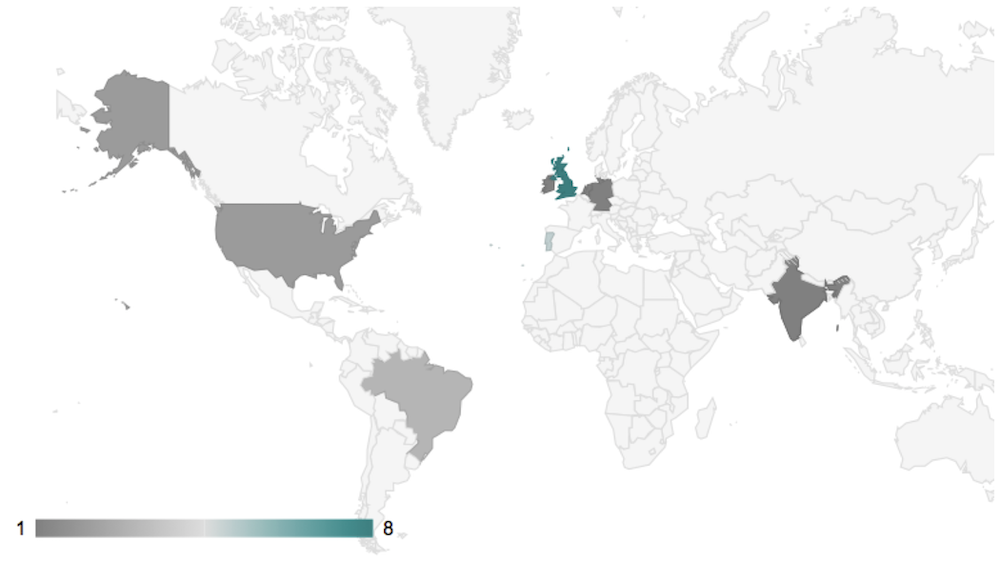}
	\caption{Publication distribution per country and continent of primary studies}
	\label{fig:DistributionContinent}
\end{figure}

Figures \ref{fig:DistributionCountry} and \ref{fig:DistributionContinent} show the distribution of studies per country. It is worth mentioning that if several countries were collaborating on the same study, we counted each country once.

\begin{figure}[h]
	\advance\leftskip 1cm
	\includegraphics[width=0.75\columnwidth]{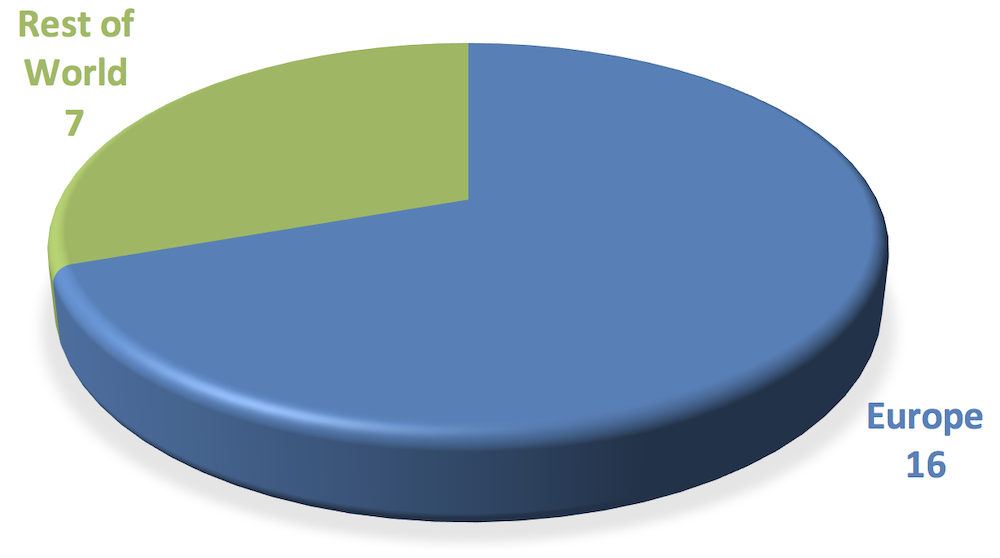}
	\caption{Publication distribution of primary studies Europe vs Rest of World}
	\label{fig:DistributionWorld}
\end{figure}

It can be argued that the UK is by far the country with the largest amount of unique proposals and collaborations. India and the USA, which are also Anglophone countries that share a number of synergies with the UK, also have a great interest in this research field. It is worth noting that Portuguese speaking countries like Portugal and Brazil also provided a great number of papers on this topic. With regard to continents, we see that 70\% of the studies were produced in Europe vs. 30\% that were produced in the rest of the world, as shown in Figure \ref{fig:DistributionWorld}. This indicates that Europe is currently a research driver within this field.

\subsubsection{Publication type }
\label{sec:Publicationtype }

In terms of type of publication, we have found that 71\% of the studies were conference papers published in papers called ``proceedings''. These types of publications are generally used to present new ideas since it takes less time to get them accepted by conferences than to get them published in journals. While some decades ago it was difficult to find conference proceedings, and researchers used to extend their conference papers into journal versions, this trend now seems to have changed owing to the current frenetic development in the era of digitalization.

As commented on in the paper ``From Conference Papers to Journal Papers: Challenges and New Ideas'' \cite{kreinovich2016conference} ``At present, for an extended version of a conference paper to be accepted in a journal, a minor extension is not sufficient because that would amount to publishing the same paper twice. To be accepted, the extended version must contain a large amount of new material. Due to this requirement, most authors who submit papers to conferences are unable to come up with follow-up journal publications''.

\begin{figure}[h]
	\includegraphics[width=\columnwidth]{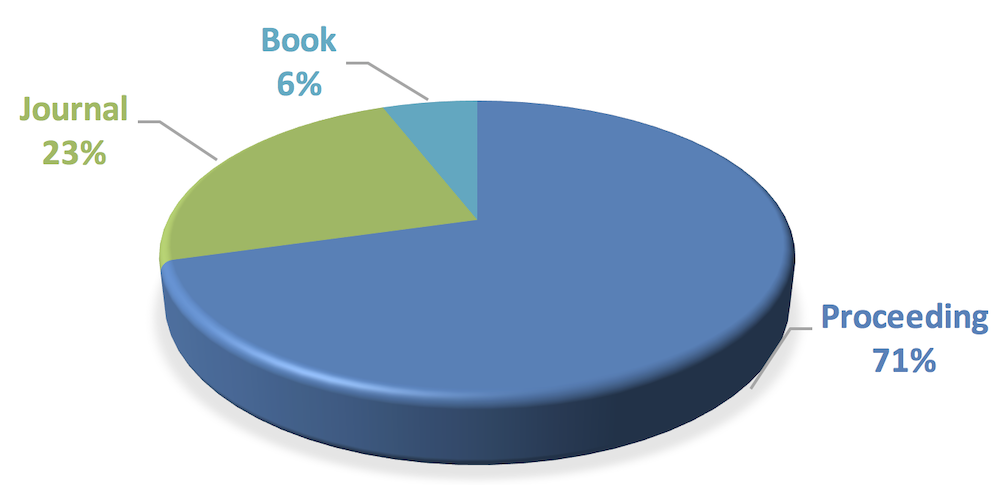}
	\caption{Publication type distribution of primary studies}
	\label{fig:PublicationType}
\end{figure}

The data we have obtained shows evidence of this behavior. Figure \ref{fig:PublicationType} summarizes the publication type distribution of primary studies.

\section{Data synthesis and results}
\label{sec:dataSynthesisResults}
This section describes the results of the present systematic review. We analyze the studies on the basis of the proposed quality assessment and research questions proposed.

\subsection{Results of the SLR} 
\label{sec:ResultsSLR}

This section discusses the answers found to respond to the research questions proposed during the planning phase of the systematic review, listed in Section \ref{sec:SLRProcess}.

This systematic review identified 17 primary studies. We found 13 unique approaches in the 17 identified. Studies [S10] [S11] [S12] and the studies [S13] [S14] [S15] belong to the same approaches, respectively. For this reason, in the review tables (\ref{tab:DetailedSyntApp} to \ref{tab:DetailedSyntSuitability}), we shall refer to S10, S11 and S12 as [S10] and S13, S14 and S15 as [S13].

\vspace{3mm}

\textbf{RQ1. What methodologies or techniques that involve creative approaches in requirements elicitation (RE) in agile software development (ASD) exist}

We have found that the selected studies present 13 different and unique proposals. According to the authors, we have found different ways of classifying the studies, and these are: ``Technique'', ``Process'', ``Framework'', ``Tool'' and ``Model''. Table \ref{tab:DetailedSyntApp} quantifies the types of proposals found in the selected studies.

\begin{table}[h]
	\caption{Types of proposals}
	\label{tab:DetailedSyntApp}
	\centering
	\begin{tabular}{lcc}
		\toprule
		\textbf{Category} & \textbf{Percentage} & \textbf{Studies} \\ \midrule
		
		\textbf{Technique} & 31\% & {\scriptsize [S3] [S5] [S10] [S13]} \\ \addlinespace
		
		\textbf{Process} & 23\% & {\scriptsize [S1] [S2] [S4]} \\ \addlinespace
		
		\textbf{Framework} & 15\% & {\scriptsize [S6] [S17]} \\ \addlinespace
		
		\textbf{Tool} & 15\% & {\scriptsize [S9] [S16]} \\ \addlinespace
		
		\textbf{Model} & 8\% & {\scriptsize [S7]} \\
		\bottomrule
	\end{tabular}
\end{table}

A technique describes a way of carrying out a specific task or procedure \cite{stappers2012bridging}. We have found four studies proposing techniques to improve the requirement elicitation process by integrating creativity. [S3] presents a technique for linking metamodels to design intent; [S5] propose to use interaction design concepts in agile development; [S10] propose a mind mapping technique for building feature models; finally, [S13] presents a goal sketching proposal for agile requirements elicitation.

A process describes an iteration of activities that must be carried out to complete a major task \cite{dowson1987iteration}. In software engineering, it can describe the whole development process or a specific sub-stage of the development life cycle. We have found three works proposing processes that integrate creativity into requirements elicitation in agile environments: ``Extended Envisioning Process'' [S1] ``Agile Design Data Modelling Process'' [S2] and ``Divingboard'' [S4].

In computer systems, a framework is often a layered structure indicating what kind of programs can or should be built and how they would interrelate \cite{vasanthi2009generic}. We have found a couple of studies presenting frameworks in the context of this review: [S6] presents the ``IBM Design Thinking Development Framework'', which combines Design Thinking and agile principles, while [S17] proposes the ``User Design Creativity Framework'', which is based on Participatory Design and creativity theories.

Software tools are programs, utilities, libraries, and other aids, such as editors, compilers, and debuggers, that can be used to develop programs \cite{aiken2002microsoft}. We have found two studies that propose tools as a means of facilitating interchange and innovation in requirements elicitation. [S9] propose an agile toolbox to share RE techniques, while [S16] propose a mobile prototyping tool to elicit requirements.

A model is a mathematical or graphical representation of a real-world situation or object \cite{aiken2002microsoft}. We just found one study [S7] proposing the ``Gamified Requirements Engineering Model'', which integrates gamification in requirements elicitation processes as a means of improving the novelty, quality and usefulness of requirements.

We were unable to find two different research groups proposing or testing the same approach. We have, however, identified a group of design methodologies or frameworks which are used as a basis on which to elaborate new approaches. To that end, they have been extended or combined with other techniques to be used for requirements elicitation in agile development. Table \ref{tab:DetailedSyntMet} shows the level of implementation of these design methodologies among the selected studies.

\begin{table}[h]
	\caption{Level of adoption of design methodologies}
	\label{tab:DetailedSyntMet}
	\centering
	\begin{tabular}{lcc}
		\toprule
		
		\textbf{Term}& \textbf{Percentage} & \textbf{Studies} \\ \midrule
		
		\textbf{Design Thinking} & 23\% & {\scriptsize [S2] [S4] [S6]}  \\ \addlinespace
		
		\textbf{Service Design Thinking} & 8\% & {\scriptsize [S3]}  \\ \addlinespace
		
		\textbf{User-Centered Design} & 15\% & {\scriptsize [S9] [S16]} \\ \addlinespace
		
		\textbf{Usage-Centered Design} & 8\% & {\scriptsize [S5]} \\ \addlinespace
		
		\textbf{Participatory Design} & 15\% & {\scriptsize [S4] [S17]} \\ 
	
		\bottomrule
	\end{tabular}
\end{table}

Design Thinking is a methodology used as a non-lineal problem-solving approach to solve complex problems and focused on the users and their needs, which aims to ensure that the developed solution meets a real user need \cite{lindberg2011design}.

Service Design Thinking is a modern interpretation of the design thinking movement in the context of providing services, mostly from businesses to customers [S3].

User-Centered Design is an approach to systems design and development that aims to make interactive systems more usable by focusing on the use of the system and applying human factors/ergonomics and usability knowledge and techniques \cite{din20109241}.

Usage-Centered Design is an approach to user interface design, focusing on user intentions and usage patterns. It analyzes users in terms of the roles they play in relation to systems and employs abstract use cases for task analysis. The term was suggested by software developer and professor Larry Constantin and Lucy Lockwood \cite{constantine2006activity}.

Finally, Participatory Design is an approach to design in which the people destined to use the system play a critical role in designing it \cite{schuler1993participatory}.

Additionally, we have identified a number of creative techniques that are used by these approaches to explore and elicit ideas. Table \ref{tab:DetailedSyntTech} shows the level of implementation of each technique among the selected studies. In the following, they are briefly defined.

\begin{table}[h]
	\caption{Level of adoption of creative techniques}
	\label{tab:DetailedSyntTech}
	\centering
	\begin{tabular}{lcc}
		\toprule
		
		\textbf{Technique} & \textbf{Percentage} & \textbf{Studies} \\ \midrule
		
		\textbf{Prototyping} & 46\% & {\scriptsize [S3] [S4] [S5] [S6] [S16] [S17]} \\ \addlinespace
		
		\textbf{Mind Mapping }& 23\% & {\scriptsize [S3] [S8] [S10]}  \\ \addlinespace
		
		\textbf{Sketching} & 23\% & {\scriptsize [S3] [S4] [S13]} \\ \addlinespace
		
		\textbf{Brainstorming} & 31\% & {\scriptsize [S1] [S4] [S5] [S6]} \\ \addlinespace
		
		\textbf{Games / Gamification} & 23\% & {\scriptsize [S3] [S6] [S7]} \\ \addlinespace
		
		\textbf{Creativity Workshops} & 23\% & {\scriptsize [S1] [S4] [S17]} \\ \addlinespace
		
		\textbf{User Stories} & 15\% & {\scriptsize [S7] [S10]} \\ \addlinespace
		
		\textbf{Random Start} & 8\% & {\scriptsize [S1]} \\ \addlinespace
		
		\textbf{PICL Technique} & 8\% & {\scriptsize [S1]} \\ \addlinespace
		
		\textbf{Storyboards} & 8\% & {\scriptsize [S6]}  \\ \addlinespace
		
		\bottomrule
	\end{tabular}
\end{table}

A prototype is a preliminary type, form, or instance of a system that serves as a model for later stages or for the final, complete version of the system \cite{ieee1990ieee}. A Mind Map is a diagram used to visualize and organize information [S8]. Sketching is a rapid drawing. Brainstorming is a process where participants from different stakeholder groups engage in informal discussion to rapidly generate as many ideas as possible without focusing on any one in particular \cite{zowghi2005requirements}. Gamification is the application of typical elements of game design elements (e.g. point scoring, competition with others, rules of play) in non-gaming contexts \cite{deterding2011game}. A User Story is a description of a feature written from the perspective of the person who needs this. It consists of a written text, conversation about it, and acceptance criteria \cite{zowghi2005requirements}. M. Michalko's Random Start is a proposal for the use of characters to force connections in order to make the generation of new requirements easier \cite{michalko2010thinkertoys}. The PCIL technique, proposed by Higgins \cite{higgins1994101}, consists basically in selecting 25 random short statements. Finally, storyboards capture stories in a visual way, combining text with visual elements to emphasise expressiveness \cite{gershon2001storytelling}.

\vspace{3mm}

\textbf{RQ2. What empirical evidence is there of the application of creative techniques for RE in ASD?}

To answer this question, we firstly looked at the type of research performed by each selected study. Table \ref{tab:DetailedSyntResearch} summarizes these data by showing the level of adoption of each research method considered. A brief discussion around each research method follows.

\begin{table}[h]
	\caption{Level of adoption of research methods}
	\label{tab:DetailedSyntResearch}
	\centering
	\begin{tabular}{lcc}
		\toprule
		
		\textbf{Type of research} & \textbf{Percentage} & \textbf{Studies} \\ \midrule
		
		\textbf{Action Research Study} & 23\% & {\scriptsize [S2] [S13] [S16]} \\ \addlinespace
		
		\textbf{Case of study} & 31\% & {\scriptsize [S4] [S10] [S11] [S17]} \\ \addlinespace
		
		\textbf{Laboratory experiment} & 23\% & {\scriptsize [S7] [S8] [S12]} \\ \addlinespace
		
		\textbf{Experience report} & 15\% & {\scriptsize [S3] [S5]} \\ \addlinespace
		
		\textbf{Field experiment} & 8\% & {\scriptsize [S1]} \\ \addlinespace
		
		\textbf{Survey} & 8\% & {\scriptsize [S6]} \\ \addlinespace
				
		\bottomrule
	\end{tabular}
\end{table}

Action Research Study is characterized by producing relevant research results because it is grounded in practical action, aimed at solving an immediate business problem \cite{baskerville1999investigating}. However, one of its disadvantages is that it is restricted to a single project and organization, and this results in a certain lack of control over variables, giving some space to different interpretations, as well as possible biases and omissions in the description and compilation of objectives \cite{cockburn2003research}.

Case studies are descriptive reports that provide in-depth analysis of a particular project or episode, usually within a single organization \cite{cockburn2003research}. As with Action Research, they are restricted to perform research in a single organization. The generalization of results is therefore difficult in some occasions.

Laboratory experiments intensively control and study a small number of variables with the objective of identifying relationships between some of those variables using quantitative and analytical techniques \cite{cockburn2003research}. The problem with this approach lies in the limitations of the identified relationship, which is somehow restricted to the context of the experiment. Real world situations are usually much more complex, including some variables that might not be considered in a controlled experiment.

Experience reports are papers written by those who participated in the completed work in a context of a real-life project \cite{cockburn2003research}. While [S3] describes an isolated experience, and provides merely anecdotal evidence, [S5] reports the results of using the proposed approach in real-world cases during some years.

Field experiments are an extension of the laboratory experiments into real-life situations. They provide greater realism than other approaches but lack of sufficient control to enable replication \cite{cockburn2003research}. Besides the documented experiment in [S1], the author mentions that the qualitative data gathered from applying the proposal in another project did support the results. He also mentions that the proposal is currently being used in other projects.

Surveys are time-stamped samples from which relationship inferences are made using quantitative analytical techniques. They help to describe real-world situations and are appropriate to generalize, but cannot ask questions regarding variables that are yet to be recognized and may present bias in participants or researchers. \cite{cockburn2003research}.

Had we looked at the empirical evidence provided by each of the studies using the criteria proposed by Kitchenham \cite{kitchenham2008evaluating} (see Table \ref{tab:EECriteria}), all the selected studies would have received a score of ``0''. The evidence provided would have therefore been considered ``poor''. The reason is that most of the studies have only tested their approaches in one or two projects.

\begin{table*}[h]
	\centering
	\caption{Empirical Evidence Criteria evaluation by Kitchenham}
	\label{tab:EECriteria}
	\resizebox{\textwidth}{!}{%
		\begin{tabular}{lllc}
			\toprule
			\textbf{Criteria}   & \textbf{Number of projects}     & \textbf{Quality}  & \textbf{Score} \\ \midrule
			\multirow{4}{*}{\begin{tabular}[c]{@{}l@{}}The size of the within-company data set, measured according to\\ the criteria presented below. Whenever a study used more than\\ one within-company data set, the average score was used\end{tabular}} & Less than 10 projects   & Poor   & 0   \\ \addlinespace
			& Between 10 and 20 projects & Fair   & 0.33 \\ \addlinespace
			& Between 21 and 40 projects & Good   & 0.66 \\ \addlinespace
			& More than 40 projects   & Excellent & 1  \\ \addlinespace
				\bottomrule
		\end{tabular}%
	}
\end{table*}

Thus, owing to the innovative nature of the research field we are studying, we have preferred to consider the evidence as a whole in order to answer RQ2, rather than evaluating each of the studies individually.

We have seen that the selected studies provide both real world and laboratory evidence and are being tested in both short and long-term projects. Given that between 10 and 20 projects have provided some sort of evidence of applying creative approaches to elicit requirements in ASD, we can conclude that the empirical evidence of applying creative approaches for requirements elicitation in ASD is moderate.

Moreover, many studies mentioned that their approaches are still being tested or would need further evidence under different circumstances.

\vspace{3mm}

\textbf{RQ3. What is known about the benefits and limitations of using creative approaches for RE in ASD? }

To respond to RQ3, we analyze the benefits and limitations mentioned in each study or those that can be inferred from the information provided. With regard to the benefits, five have been commonly identified in selected studies, as detailed in Table \ref{tab:DetailedSyntBenefits}.

\begin{table}[h]
	\caption{Identified benefits}
	\label{tab:DetailedSyntBenefits}
	\centering
	\begin{tabular}{p{3.2cm}cp{2.5cm}}
		\toprule
		
		\textbf{Benefit} & \textbf{Percentage} & \textbf{Studies} \\ \midrule
		
		Facilitate the understanding of the problem & 69\% & {\scriptsize [S3] [S4] [S5] [S6] [S9] [S10] [S13] [S16] [S17]}  \\ \addlinespace
		
		Better meet business needs & 54\% & {\scriptsize [S2] [S5] [S6] [S9] [S10] [S13] [S16]}  \\ \addlinespace

		Improvement of stakeholder engagement & 31\% & {\scriptsize [S5] [S10]] [S16] [S17]} \\ \addlinespace
		
		Improvement in communication and ideas interchange & 31\% & {\scriptsize [S2] [S4] [S16] [S17]}  \\ \addlinespace
		
		Improvement in novelty, quality and usefulness of requirement & 23\% & {\scriptsize [S1] [S7] [S17]}  \\ \addlinespace		
		
		\bottomrule
	\end{tabular}
\end{table}

It has been observed that most of the selected studies aim at dealing with problem by understanding issues. By facilitating the understanding of the problem, requirements elicitation becomes more manageable and oriented exactly towards user needs. Furthermore, most of the selected studies showed that empowering creativity in requirements elicitation results in the development of products that better meet user demands. Likewise, a certain number of the selected studies present approaches that facilitate and improve communication between stakeholders, thus facilitating the exchange of ideas and productive collaboration. The same applies to techniques or approaches that intensively encourage stakeholder participation. Finally, only three studies specifically mentioning improvement in novelty and quality of requirements have been found. This is probably due to the fact that evaluating innovation or quality is not an easy task.

Regarding the limitations of adopting creative techniques for requirements elicitation in ASD, it is worth mentioning that the selected studies rarely discussed the main issues of their proposals. Nevertheless, Table \ref{tab:DetailedSyntLimitations} summarizes the limitations that we were able to identify.

\begin{table}[h]
	\caption{Identified limitations}
	\label{tab:DetailedSyntLimitations}
	\centering
	\begin{tabular}{p{3.2cm}cp{2.5cm}}
		\toprule
		
		\textbf{Limitations} & \textbf{Percentages} & \textbf{Studies} \\ \midrule
		
		Requirements obtained are seen as less useful at the very early stages & 8\% & {\scriptsize [S1]} \\ \addlinespace
		
		Users should be open-minded, willing to collaborate and patient & 31\% & {\scriptsize [S4] [S5] [S16] [S17]} \\ \addlinespace
		
		Reorganization of teams might be needed & 8\% & {\scriptsize [S6]}  \\ \addlinespace
		
		Training in novel techniques would be needed & 69\% & {\scriptsize [S1] [S2] [S3] [S4] [S5] [S6] [S8] [S10] [S13]}  \\ \addlinespace
		
		\bottomrule
	\end{tabular}
\end{table}

First of all, even though they do not mention this explicitly, those studies presenting novel techniques require some sort of training. Therefore, they should consider this limitation in terms of time (needed to train the stakeholders) and other resources. Another of the identified limitations is that users and organizations must be open-minded, willing to collaborate and patient. As a matter of fact, in order to apply these approaches, users are required to collaborate intensively, what may produce fatigue. For instance, [S6] specifically stated that teams and departments within the organization might need to be reorganized. This is clearly a limitation because many organizations won't allow or facilitate this type of proceedings. Finally, [S1] mentioned that during the initial stages, requirements obtained by means of creative techniques were seen as less useful. It is worth noting, however, that their usefulness became much more evident in later stages after incubation.

\vspace{3mm}

\textbf{RQ4. In what sort of organizations or projects using ASD could the use of these creative approaches be most suitable? }

During the execution of the present systematic review we noticed that RQ4 might be too ambitious given the innovative nature of the area. Since the selected studies present only one or two pieces of evidence, determining the suitability of the proposals for certain projects or organizations is quite complex. All this given, Table \ref{tab:DetailedSyntSuitability} only shows the information collected in the selected studies about the projects in which these proposals have been applied. This does not necessarily mean that one approach is suitable only for these types of projects.

We have proceeded with the following strategy: based on the information gathered from each study, we identified and defined three project categories that could help us respond to this research question:

	\begin{enumerate}

		\item \textit{Organization size}: small (startup or projects with less than 10 participants) vs. large organizations (10 or more participants).

		\item \textit{Duration of the project}: short-term (less than one year) vs. long-term (more than one year).

		\item \textit{Type of development}: mobile applications, website development, enterprise architecture and web application.

	\end{enumerate}

After reviewing the studies according to these criteria, we obtained the results summarized in Table \ref{tab:DetailedSyntSuitability}. Note that the sum of percentages within one category (e.g. large + small organizations) is not necessarily 100\%. The reason for this is that we evaluated the categories individually. One specific approach might have been successfully implemented in both large and small organizations.

\begin{table}[h]

	\caption{Detailed synthesis of suitability}

	\label{tab:DetailedSyntSuitability}

	\begin{tabular}{p{1.4cm}lp{2.3cm}}

		\toprule

		\textbf{Category} & \textbf{Percentage} & \textbf{Studies} \\ \midrule

		\textbf{Organization size} & Large 54\% & {\scriptsize [S1] [S2] [S4] [S5] [S6] [S10] [S13]}  \\ \addlinespace

		& Small 15\% & {\scriptsize [S10] [S13]}   \\ \addlinespace

		& Small and Large 15\% & {\scriptsize [S10] [S13]}  \\ \addlinespace \addlinespace

		\textbf{Duration} & Short: 15\% & {\scriptsize [S13] [S16]}  \\ \addlinespace

		& Long: 31\% & {\scriptsize [S2] [S4] [S5] [S6]}   \\ \addlinespace \addlinespace

		\textbf{Product type} & Mobile applications 31\% & {\scriptsize [S6] [S10] [S16] [S17]}  \\ \addlinespace

		& 31\% Enterprise architecture & {\scriptsize [S3] [S6] [S10] [S13]}  \\ \addlinespace

		& 31\% Web applications & {\scriptsize [S4] [S6] [S10] [S13]}  \\ \addlinespace

		& 8\% Website development & {\scriptsize [S1]} \\ \addlinespace

		\bottomrule

	\end{tabular}

\end{table}

According to the data obtained, 54\% of the approaches that mentioned the sort of organization in which their approaches were being tested or could be used referenced large organizations. On the contrary, only 15\% of the approaches were deployed in small organizations. This might be related to the fact that most of the reviewed approaches require the final user to be intensively involved. Quite often, this implies an extra cost which small organizations cannot always afford. Finally, six studies (two proposals, [S10] and [S13]) were found suitable for both, small and large organizations.

Regarding project length, most of the proposals were tested during a reduced amount of time, which mostly involved just the phase of requirements gathering. It was therefore difficult to assess the proposals in terms of project length. From the information that we could extract or infer we found 2 approaches, ([S13] and [S16]), being used or suitable for short projects. On the contrary, we found 4 studies (31\%) being used in long-term development projects. We found one study [S6] mentioning that the approach would not be suitable for short-term projects as well.

Regarding project nature, we found 4 studies applying their proposal to: the development of mobile apps, business management systems and web apps. Finally, one of the studies was applied in the development of a web-site.

\vspace{3mm}

\textbf{RQ5. Which agile methodologies are being used to integrate creative techniques into agile software development RE processes?}

Very few studies mention a particular agile methodology being used to integrate these approaches. We found, however, four proposals mentioning some kind of agile model-driven development [S10], [S11], [S12], [S17]; two more works specifically mentioning the use of Scrum or employing traditional Scrum elements like sprints and product backlog [S1], [S8]; and finally, one study using Extreme programming [S5].

In general, most of the works prefer to propose approaches that could be used with different agile methodologies instead of restricting their application to a particular method or technique.

\subsection{Final Remarks}
\label{sec:FinalRemarks}

This systematic review has provided an overview of the existing approaches that empower creativity in requirements elicitation within agile software development. 

We have found that agile methodologies like Scrum, Extreme Programming or methodologies based on rapid modelling are preferred to enhance creativity in requirements elicitation. 

Regarding the empirical evidence, due to the novelty of the research field, we have not found strong evidence. However, considering the results that we have obtained with this systematic review, empowering creativity into requirements elicitation is not a theoretical idea any longer, but a mixture of techniques, tools and processes that have been and are being successfully tested in the industry. 

Due to the novelty and variety of approaches, which have been mostly tested in only one or two projects so far, it is complicated to conclude in what sort of projects these techniques might be more suitable. However, it has been observed that highly interactive nature development projects like mobile or web applications are good candidates to make use of these creative approaches. 

Probably also due to the novelty of the area, the reviewed works do not consider creativity from a scientific point of view. For instance, none of the reviewed works consider the four C model of creativity from Kaufman and Beghetto \cite{kaufman2009beyond}. Current landscape leaves space to more scientific approaches from agile practitioners when aiming at leveraging creativity in requirements elicitation.

Finally, we have found that, although creativity is an important ingredient to bring about innovation, it is not always sufficient to generate novel requirements. It needs to be accompanied by user engagement and a specific context where proper conditions like flexibility, time or resources have to be met.

\section{Conclusions and further work}
\label{sec:conclusions}
This Section summarizes the conclusions and future work of this article. To accomplish this, we analyze the achievement of objectives defined in Section \ref{sec:introduction} and the future work that has been identified.

\subsection{Achievement of objectives}
\label{sec:AchievementsOjectives}

At the beginning of this study, in Section \ref{sec:introduction} we defined several sub-goals that were needed to achieve the main goal: \textit{to study the state-of the art in the approaches that strength creativity in requirements elicitation within agile software development}. 

We now analyze how and to what extent these goals have been met:

\vspace{3mm}

\textbf{Sub-Goal 1} -\textit{ Research on background studies}

In order to achieve this goal, we have carried out a research on systematic literature reviews, with a special focus on systematic literature reviews for software engineering. We have also researched the topic of requirements elicitation and agile methodologies (providing details of Scrum, Extreme Programming and Kanban). Finally, we have investigated creativity, firstly from a theoretical point of view, and secondly as regards the application of creativity in design (with a special focus on Design Thinking) and requirements elicitation. All this research has been documented in Section \ref{sec:researchContext}.

\vspace{3mm}

\textbf{Sub-Goal 2} - \textit{Definition of a criterion that will serve to select and evaluate relevant studies}

This goal has been fulfilled in Section \ref{sec:SLRProcess} during the planning stage of the systematic review. Here, we defined a quality assessment criterion in terms of rigor, credibility and relevance. We also defined a classification criterion that would help us to better organize the proposals and summarize the results.

\vspace{3mm}

\textbf{Sub-Goal 3} - \textit{Execution of a systematic literature review}

To fulfill this goal, we executed the search strategy that was defined during the planning stage and identified 1451 articles, of which 17 papers were included in the review as primary study papers. In Section \ref{sec:SLRProcess}, we provide a detailed description and present an overview of the selected studies.

\vspace{3mm}

\textbf{Sub-Goal 4} - \textit{Review and summary of the selected studies and identification of trends}

Finally, in order to accomplish this goal we carried out a detailed review of each of the selected studies, as documented in Section \ref{sec:dataExtraction}. Later, in Section \ref{sec:dataSynthesisResults} we analyzed and classified all gathered information .

\subsection{Future work}
\label{sec:FutureWork}

Having finalized the present study, some improvements and extensions were identified.

\begin{itemize}
	
	\item \textit{Further systematic reviews}: since this is a very new area of study, further systematic reviews are needed to identify works that have been published after this systematic review was executed. It would also be interesting to employ another combination of keywords, such as those related to innovation and user engagement. To that end, new systematic reviews could be performed using the present review as a basis.
	\item \textit{Improvements related to the insights from the industry}: in order to attain a detailed overview of the current situation in industry, a complete questionnaire should be created. It could be used to analyze the specific circumstances of professionals and projects in terms of location and resources, as well as to contrast the findings of this research with the current situation of the industry.
	\item \textit{Extension of creativity approaches to other stages of development}: we have studied creativity during the design and requirements elicitation phases. It would be interesting to investigate the employment of creativity in other areas of development, such as testing or integration. In order to do so, new reviews should be performed so as to assess whether there are proposals aiming at leveraging creativity at other stages of the development process.
	\item \textit{Combination of proposals}: it would also be interesting to analyze a possible combination of techniques such as creativity workshops with mind mapping or sketching. For this to happen, a development team and some experts on psychology, soft-skills and the like are needed to design some kind of experiment.
	\item \textit{More empirical evidence and elaborated proposals}: since this is a very new and immature area, there is a need to apply these and other new approaches in more real scenarios. In this sense, it might be the case that we should not limit the scope to the development of software projects.
\end{itemize}

\begin{acknowledgements}
\sloppy
This research has been partially funded by the Regional Government of Madrid under the FORTE-CM (S2018/TCS-4314) project and the MADRID (TIN2017-88557-R) project, financed by the Spanish Ministry of Economy and Business.
\end{acknowledgements}


\bibliographystyle{spmpsci}      
\bibliography{JSS_SLR}  
\clearpage


\onecolumn
\begin{appendices}

	\chapter{\textbf{Appendix I - List of studies selected for the review (Primary Studies)}}

	\begin{itemize}
		
		\item[\textbf{S1}] Hollis, B., \& Maiden, N. (2013). Extending agile processes with creativity techniques. IEEE software, 30(5), 78-84 \vspace{3mm}
		
		\item[\textbf{S2}]  O’Driscoll, K. (2016). The agile data modelling \& design thinking approach to information system requirements analysis. Journal of Decision Systems, 25(sup1), 632-638. \vspace{3mm}
		
		\item[\textbf{S3}]  Gamble, M. T. (2016). Can metamodels link development to design intent?. In Proceedings of the 1st International Workshop on Bringing Architectural Design Thinking into Developers' Daily Activities, 14-17  \vspace{3mm}
		
		\item[\textbf{S4}] Newman, P., Ferrario, M. A., Simm, W., Forshaw, S., Friday, A., \& Whittle, J. (2015). The role of design thinking and physical prototyping in social software engineering. 37th International Conference on Software Engineering, 2, 487-496 \vspace{3mm}
		
		\item[\textbf{S5}] Patton, J. (2002). Hitting the target: adding interaction design to agile software development. In OOPSLA 2002 Practitioners Reports, 1-ff \vspace{3mm}
		
		\item[\textbf{S6}] Lucena, P., Braz, A., Chicoria, A., \& Tizzei, L. (2016). IBM Design Thinking Software Development Framework. In Brazilian Workshop on Agile Methods, 98-109 \vspace{3mm}
		
		\item[\textbf{S7}] Lombriser, P., Dalpiaz, F., Lucassen, G., \& Brinkkemper, S. (2016). Gamified requirements engineering: model and experimentation. In International Working Conference on Requirements Engineering,171-187 \vspace{3mm}
		
		\item[\textbf{S8}] Mahmud, I., \& Veneziano, V. (2011). Mind-mapping: An effective technique to facilitate requirements engineering in agile software development. In Computer and Information Technology (ICCIT), 2011, 157-162 \vspace{3mm}
		
		\item[\textbf{S9}] Maiden, N. (2011). What Time Is It, Eccles?. IEEE software, 28(4), 84-85 \vspace{3mm}
		
		\item[\textbf{S10}] Wanderley, F., da Silveira, D. S., Araujo, J., \& Lencastre, M. (2012). Generating feature model from creative requirements using model driven design. In Proceedings of the 16th International Software Product Line Conference, 2, 18-25 \vspace{3mm}
		
		\item[\textbf{S11}] Wanderley, F., \& Araujo, J. (2013). Generating goal-oriented models from creative requirements using model driven engineering. In Model-Driven Requirements Engineering (MoDRE), 2013 International Workshop on,1-9 \vspace{3mm}
		
		\item[\textbf{S12}] Wanderley, F., Silveira, D., Araujo, J., Moreira, A., \& Guerra, E. (2014). Experimental evaluation of conceptual modelling through mind maps and model driven engineering. In ICCSA 2014, 200-214 \vspace{3mm}
		
		\item[\textbf{S13}] Boness, K., \& Harrison, R. (2007). Goal sketching: Towards agile requirements engineering. In Software Engineering Advances, ICSEA 2007, 71-71 \vspace{3mm}
		
		\item[\textbf{S14}] Boness, K., \& Harrison, R. (2008). Goal sketching with activity diagrams. In Software Engineering Advances, 2008. ICSEA'08. The Third International Conference on, 277-283 \vspace{3mm}
		
		\item[\textbf{S15}] Boness, K., Harrison, R., \& Liu, K. (2008). Goal sketching: An agile approach to clarifying requirements. International Journal on Advances in Software, IARIA, 1(1) \vspace{3mm}
		
		\item[\textbf{S16}] Hastreiter, I., Krause, S., Schneidermeier, T., \& Wolff, C. (2014). Developing UX for Collaborative Mobile Prototyping. In International Conference of Design, User Experience, and Usability, 104-114 \vspace{3mm}
		
		\item[\textbf{S17}] Sulmon, N., Derboven, J., Zaman, B., \& Montero, M. (2012). Mapping Participatory Design Methods to the Cognitive Process of Creativity to Facilitate Requirements Engineering. Information Systems Research and Exploring Social Artefacts: Approaches and Methodologies, 221-241 
		
	\end{itemize}

\end{appendices}

\end{document}